\documentclass{article}
\usepackage{amsthm}

\newtheorem{proposition}{Proposition}
\newtheorem{corollary}{Corollary}

\usepackage{PRIMEarxiv}

\usepackage[utf8]{inputenc} 
\usepackage[T1]{fontenc}    
\usepackage{hyperref}       
\usepackage{url}            
\usepackage{booktabs}       
\usepackage{amsfonts}       
\usepackage{nicefrac}       
\usepackage{microtype}      
\usepackage{lipsum}
\usepackage{fancyhdr}       
\usepackage{graphicx}       
\graphicspath{{media/}}     
\usepackage[super]{nth}
\usepackage{natbib}
\usepackage{amsmath}
\usepackage{moreverb}
\usepackage{todonotes}
\usepackage{amsmath}
\usepackage{subcaption}
\usepackage{multirow}
\usepackage{csquotes}
\usepackage{xspace}
\usepackage{listings}
\usepackage[super]{nth}
\usepackage{appendix}
\usepackage{comment}
\usepackage{float}

\newcommand*{\aias}{AI-augmented system}

\title{Risks of ignoring uncertainty propagation in AI-augmented security pipelines
\thanks{\textit{\underline{Citation}}: 
\textbf{Mezzi, E., Papotti, A., Massacci, F., \& Tuma, K. (2025). Risks of ignoring uncertainty propagation in AI-augmented security pipelines. Risk Analysis, 1--21. DOI: \url{https://doi.org/10.1111/risa.70059}.}} 
}

\author{
  Emanuele Mezzi \\
  Vrije Universiteit Amsterdam \\
  \texttt{e.mezzi@vu.nl} \\
   \And
  Aurora Papotti \\
  Vrije Universiteit Amsterdam \\
  \texttt{a.papotti@vu.nl} \\
   \And
   Fabio Massacci \\
   Vrije Universiteit Amsterdam \\
   University of Trento \\ 
   fabio.massacci@ieee.org \\
  \And
  Katja Tuma \\
  Eindhoven University of Technology \\  
  \texttt{k.tuma@tue.nl} \\
}

\begin{document}
\maketitle

\begin{abstract}
The use of AI technologies is being integrated into the secure development of software-based systems, with an increasing trend of composing AI-based subsystems (with uncertain levels of performance) into automated pipelines. This presents a fundamental research challenge and seriously threatens safety-critical domains. Despite the existing knowledge about uncertainty in risk analysis, no previous work has estimated the uncertainty of \aias s given the propagation of errors in the pipeline. We provide the formal underpinnings for capturing uncertainty propagation, develop a simulator to quantify uncertainty, and evaluate the simulation of propagating errors with one case study. We discuss the generalizability of our approach and its limitations and present recommendations for evaluation policies concerning AI systems. Future work includes extending the approach by relaxing the remaining assumptions and by experimenting with a real system.
\end{abstract}


\section{Introduction}
\label{sec:introduction}
Due to the increasing availability of data, AI technologies have spread and are being used in almost every computing system, including in safety-critical domains (Perez-Cerrolaza et al., \citeyear{perez2024artificial}).
Although the use of \aias s comes with new promises of improved performance, it also introduces significant risks and challenges (Cox Jr, \citeyear{cox2020answerable}; Nateghi \& Aven, \citeyear{nateghi2021risk}). A major challenge in using AI for risk analysis is conveying to decision-makers the uncertainty inherent to predictions of models, because it clashes with the common practice in the realm of AI to communicate uncertainty with point estimates or ignoring it completely (Guikema, \citeyear{guikema2020artificial}).  

With the rise of open-source software development and large-scale cloud deployment, more security risk decision-making is automated by running sequences of AI-augmented analyses like automated program repair (APR) (Long et al., \citeyear{long2017automatic}; Ye et al., \citeyear{ye2021automated}; Li et al., \citeyear{li2022dear}; Xia \& Zhang, \citeyear{xia2022less}; Fu et al., \citeyear{fu2024aibughunter}).
The use of AI in automated security pipelines, where the first classifier detects a vulnerability and the second tool fixes it, is now becoming more common (Bui et al., \citeyear{bui2024apr4vul}), bringing about a fundamental research challenge:

\smallbreak
\textit{\textbf{Propagating uncertainty} is a new major challenge for assessing the \textbf{risk} of automated security pipelines.}
\smallbreak

This foundational problem has already manifested in security pipelines with no AI-based computation. To illustrate this problem we consider four studies: verifying the presence of code smells (Tufano et al., \citeyear{tufano2017and}), generalizing the SZZ algorithm to identify the past versions of software affected by a vulnerability (Dashevskyi et al., \citeyear{dashevskyi2018screening}), identifying vulnerabilities in Java libraries (Kula et al., \citeyear{kula2018developers}), and finding how vulnerable Android libraries could be automatically updated (Derr, \citeyear{derr2017keep}).

A few years later, Pashchencko et al. (\citeyear{pashchenko2020vuln4real}) showed that the results by Kula et al. (\citeyear{kula2018developers}) are incorrect, and Huang et al. (\citeyear{huang2019up}) found that the claims by Derr et al. (\citeyear{derr2017keep}) are incorrect, both to a large extent. We argue that the reason for this mishap is foundational. All these studies share the impossibility of running manual validation and do not report the uncertainty of their outcomes. 
The proposed solutions process huge inputs (e.g., 246K commits in Dashevskyi et al. (\citeyear{dashevskyi2018screening})) so they need an automated tool, with an error rate, to decide whether each sample satisfies the property of interest.

With the appearance of new AI-based approaches, such as SeqTrans (Chi et al., \citeyear{chi2022seqtrans}), it is becoming imperative to investigate this problem now, before it is too late and {\aias s} without global measures of risk become weaved into the automated pipelines in organizations. 

To address these issues,  we focus on understanding the uncertainty due to error propagation in AI Augmented Systems. Among the pipeline, each component may be a potential source of error that leads to an underestimation or overestimation of the actual effectiveness of the proposed solution. Therefore, we formulate the overarching research question:

\smallbreak
\noindent
\textbf{RQ: }\emph{How to estimate the total error (or success rate) of the \aias, given the propagating errors of
the classifiers in the pipeline?}
\smallbreak

If analytical models for the classifiers and the fixer components existed, it could be possible to use the error propagation models used for calculus (Benke et al., \citeyear{benke2018error}). Unfortunately, analytical models of the recall and precision of these tools are extremely rare, therefore, we must resort to the much coarse-grained approximation with probability bound analysis (PBA) (Iskandar, \citeyear{iskandar2021probability}). 

\subsection{Contributions}
\label{sec:contributions}
We provide the formal underpinnings for capturing uncertainty propagation in AI-augmented APR pipelines. In addition, we develop a simulator to quantify the effects that propagating uncertainty has in automated APR tools (such as the one presented in Figure \ref{fig:pipeline}). We evaluate the simulator and present one case study in which we calculate the effects of uncertainty regarding the proposed solution. We provide the code in a GitHub repository (Mezzi \& Papotti, \citeyear{softwareRepo}). Finally, given our findings, we discuss recommendations for the evaluation policies concerning AI systems. 

\section{Background and related work}
\label{sec:state_of_art}
As background, we illustrate the composition of AI-augmented APR tools and present related work on uncertainty quantification in AI and the applications of AI to vulnerability detection and APR.

\def\fixrate{\ensuremath{f_R}}
\def\fixaias{\ensuremath{f(aias)}}
\def\recall{\ensuremath{rec}}
\def\precision{\ensuremath{prec}}
\def\prevalence{\ensuremath{P_R}}
\def\prevalenceaias{\ensuremath{P_R(aias)}}
\def\positives{\ensuremath{Pos}}
\def\negatives{\ensuremath{Neg}}
\def\total{\ensuremath{N}}
\def\specificity{\ensuremath{spec}}
\def\sem{\ensuremath{\sigma_{\bar{X}}}}

\subsection{AI-augmented systems}
\label{sec:ai-augm-sys}
Figure \ref{fig:pipeline}, shows the simplest example of composed \aias \@\xspace in the area of APR. It is composed by \textit{(i)} a classifier, which labels code samples as \emph{Good} or \emph{Bad} by detecting which sample is not vulnerable and which is vulnerable, \textit{(ii)} a fixer tool to transform \emph{Bad} samples into \emph{Good} samples, \textit{(iii)} and the second classifier, which can be either equal to the first or different, which analyzes the samples modified by the fixer to check whether they have been successfully repaired. The outcome of the final step is what we call \emph{claimed success rate} or \emph{fix rate}, representing the ratio of fixed vulnerable samples concerning the total number of vulnerable samples. Here we list each step executed by the AI-augmented APR tool and the possible errors propagating from it: 

\begin{figure*}
    \centering
    \includegraphics[width=\textwidth]{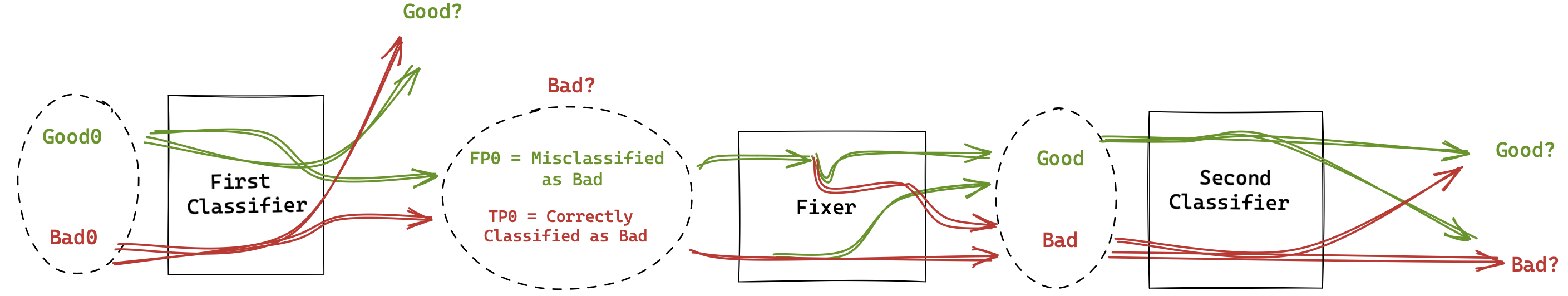}
    \caption{Illustration of an AI-augmented system which performs vulnerability detection and program repair. The first classifier receives in input the code samples and determines which are the positive samples to be sent to the fixer to be repaired. The fixer, based on its effectiveness, tries to repair them. The second classifier checks whether the fixing is correct. We can also observe the errors made by each component of the pipeline. The first classifier can wrongly classify samples as positive when they are negative. The fixer can fail in repairing positive samples, and the second classifier can misclassify the samples received and that were modified by the fixer.}
    \label{fig:pipeline}
\end{figure*} 

\begin{itemize}
    \item Step one: the first classifier analyzes the code samples, and labels each of them as \emph{Good} or \emph{Bad}. If a code sample presents features not encoded in the distribution learned by the classifier, misclassification is probable, and thus the possibility that a \emph{Good} code sample is misclassified as \emph{Bad} or vice-versa. 

    \item Step two: the fixer tries to fix every \emph{Bad} code sample, transforming it into a \emph{Fixed} code sample. Here the possibility of error lies in the fixer's performance. 

    \item Step three: the second classifier analyzes the \emph{Fixed} code samples. This is the outcome of the entire system. The second classifier performs a final analysis to detect which applications have not been successfully fixed by the fixer. The possibility of errors lies in the same conditions defined for the first classifier.
\end{itemize}

In our research, we focus on the errors of the first and second classifiers by modeling and propagating the uncertainty which characterizes the classifier's capacity to spot vulnerable code. The classifier's capacity to detect vulnerable code is measured by the Recall (\recall) or True Positive Rate ($TPR$), which is the ratio between the true positives and all the positive samples. We formally define the Recall in Section \ref{sec:methodology}. 

\subsection{Uncertainty quantification in AI}
Hüllermeier and Waegeman (\citeyear{hullermeier2021aleatoric}), 
highlight two macro-categories of methods employed to quantify and manage uncertainty in Machine Learning (ML). The first discerns between frequentist-inspired and Bayesian-inspired quantification methods. The second considers the distinction between uncertainty quantification and set-values prediction. Uncertainty quantification methods allow the model to output the prediction and the paired level of certainty, while the set-value methods consist of pre-defining a desired level of certainty and producing a set of candidates that comply with it. 

Abdar et al. (\citeyear{abdar2021review}) focus their analysis on Deep Learning (DL). Bayesian-inspired methods and ensemble methods represent two of the major categories to represent uncertainty in DL. Through Bayesian methods, the DL model samples its parameters from a learnt posterior distribution, allowing the model to avoid fixed parameters and allowing us to inspect the variance and uncover the uncertainty which surrounds the model predictions. The most common Bayesian-inspired technique is the Monte Carlo (MC) dropout. Ensemble methods combine different predictions from different deterministic predictors. Although they were not introduced in the first instance to explicitly handle uncertainties, they give an intuitive way of representing the model uncertainty on a prediction by evaluating the variety among the predictors (Gawlikowski et al., \citeyear{gawlikowski2023survey}). 

\textbf{Key Observation 1.} Extensive research was performed in the field of uncertainty quantification in AI, which brought the development of a variety of methods. However, these approaches focus on uncertainty quantification of isolated models without accounting for how uncertainty characterizing a model's output can propagate and impact subsequent system components when the model is part of an AI-augmented pipeline and its output constitutes the input to other models.

\subsection{AI in vulnerability detection}
\label{sec:AI-in-vuln-dect}
Vulnerability detection is a crucial step in risk analysis of software systems and includes running automated tools scanning parts of the system to prevent future exploitation. 
Given its potential, experts integrated AI into their vulnerability detection systems, to scale them and make them more flexible to new threats.

One of the approaches to perform vulnerability detection is obtained by applying Natural Language Processing. In their approach, Hou et al. (\citeyear{hou2022vulnerability}), represent the code in the form of a syntax tree and input it to a Transformer model, which leverages the attention mechanism to improve the probability of detecting vulnerabilities. Akter et al. (\citeyear{akter2022automated}), create embeddings using GloVe (Pennington et al., \citeyear{pennington2014glove}) and fastText (Joulin et al., \citeyear{joulin2016fasttext}), word embedding methods which aim to capture the relations between words. Then, they use LSTM and Quantum LSTM models to perform vulnerability detection, showing lower execution time and higher accuracy, precision, and recall for the Quantum LSTM. 

Another line of research excludes Natural Language Processing or embeds it with graph approaches. Yang et al. (\citeyear{yang2022vdhgt}), propose a new code representation method called vulnerability dependence representation graph, allowing the embedding of the data dependence of the variables in the statements and the control structures corresponding to the statements. Moreover, they propose a graph learning network based on a heterogeneous graph transformer, which can automatically learn the importance of contextual sentences for vulnerable sentences. They carry out experiments on the SARD dataset (NIST, \citeyear{sarddataset}) with an improvement in performance between 4.1\% and 62.7\%. Fan et al. (\citeyear{fan2023vdotr}), propose a circle gated graph neural network (CGGNN) that receives an input tensor structure used to represent information of code. CGGNN possess the capacity to perform heterogeneous graph information fusion more directly and effectively which allows the researchers to reach a higher accuracy precision and recall compared to the TensorGCN (Liu et al., \citeyear{liu2020tensor}) and Devign (Zhou et al., \citeyear{zhou2019devign}) methods. 

Finally, Zhang et al. (\citeyear{zhang2023vulgai}) propose VulGAI to overcome the limitations posed by the training time in graph neural network models. They base their methods on graphs and images and unroll their approach in four phases: the graph generation from the code, the node embedding and the image generation from the node embedding. Then, vulnerability detection through convolutional neural networks (CNN) is applied. 

VulGAI was tested on 40657 functions, outperforming other methods such as VulDePecker, SySeVR, Devign, VulCNN, and mVulPreter. Furthermore, VulGAI showed high accuracy, recall, and f1-score, improving by 3.9 times the detection time of VulCNN.

\textbf{Key Observation 2.} Extensive research and different approaches have been tested in the past, with a high level of performance. However, previous work does not quantify (or communicate) the uncertainty regarding the performance of the proposed methods, and yet, the overestimated performance of the vulnerability detection model could affect the entire pipeline performance.

\subsection{APR and composed pipelines}
The step which follows automatic vulnerability detection through AI is the application of AI to automatic code fixing. 

\subsubsection{Code fixers} 
Li et al. (\citeyear{li2022dear}) propose DEAR, a DL approach which supports fixing general bugs. Experiments run on three selected datasets: Defects4J (395 bugs), BigFix (+26k bugs), and CPatMiner (+44k bugs) show that the DEAR approach outperforms existing baselines. Chi et al. (\citeyear{chi2022seqtrans}) leverage Neural Machine Translation (NMT) techniques, to provide a novel approach called SeqTrans to exploit historical vulnerability fixes to automatically fix the source code. Xia \& Zhang (\citeyear{xia2022less}), propose AlphaRepair, which directly leverages large pre-trained code models for APR without any fine-tuning/retraining on historical bug fixes. 

\subsubsection{Composed pipelines}
\label{sec:apr_composed_pipelines}
AIBUGHUNTER combines vulnerability detection and code repair. The pipeline is implemented by Fu et al. (\citeyear{fu2024aibughunter}), combining LineVul (Fu \& Tantithamthavorn, \citeyear{fu2022linevul}) and VulRepair (Fu et al., \citeyear{fu2022vulrepair}), two software implemented by the same author. Yang et al. (\citeyear{yang2020applying}) propose a DL approach based on autoencoders and CNNs, automating bug localization and repairs. Another example of a complete pipeline combining vulnerability detection and code repair is HERCULES, which employs ML to fix code (Saha et al., \citeyear{saha2019harnessing}). Liu et al. (\citeyear{liu2021critical}), evaluate the effect of fault localization by introducing the metric \emph{fault localization sensitiveness (Sens)} and analyzing 11 APR tools. \emph{Sens} is calculated with the ratio of plausibly fixed bugs by modifying the code on non-buggy positions, and the percentage of bugs which could be correctly fixed when the exact bug positions are available but cannot be correctly fixed by the APR tool with its normal fault localization configuration. This metric, to the best of our knowledge, is the first to quantify the impact of the vulnerability detector capability on the overall pipeline. Nevertheless, it does not provide an interval to describe the best and worst pipeline performance, and thus the quantification of the risk in terms of the percentage of errors which the pipeline will overlook when it is employed.

\textbf{Key observation 3.} Recently, substantial research has appeared regarding the automation of vulnerability fixing by using ML. These advances are important and could help to manage the manual effort spent on sieving through tool warnings. However, to the best of our knowledge, the propagation of errors (or final uncertainty of the result) has not been investigated in such automated pipelines.

\section{Pipeline formalization}

\label{sec:methodology}

In this section, we present the formal basis for our simulator. To simplify the analysis, we make the following assumptions in our model:

\begin{itemize}
    \item \emph{No breaking}: 
    We assume that the fixer will never turn a true \emph{Good} sample that is classified as \emph{Bad} into a \emph{Bad} sample.
    \item \emph{No degradation}: 
    We assume that all elements that are fixed, cannot be distinguished from \emph{Good} elements from the beginning. In other words, the performance of the second classifier does not degrade with the fix. 
    \item \emph{Constant prevalence rate}: We initially assume that the prevalence rate $\prevalence$ which defines the number of positive samples in the dataset is the same for both the training and the test dataset. We relax this assumption in Section \ref{sec:simulation_two}.
\end{itemize}

\subsection{Identify the classifier metrics} 
\label{sec:sec:classifiermetrics}

To evaluate the performance of the \aias\ we use the metrics which are typically used to report the performance of a classifier: True Positive Rate ($TPR$) or Recall (\recall), precision (\precision), and False Alert Rate ($FAR$) or False Positive Rate ($FPR$), which we use interchangeably throughout this manuscript. We also use the prevalence rate (\prevalence) of the positive elements (\positives) among the total number of objects (\total) in the domain of interest. The prevalence rate is not typically known, so we will assume it to be a parameter whose effects need to be explored by simulation. Specificity is rarely cited in publications using AI models and its absence makes it difficult to reverse engineer the True Negatives.  

\begin{align}
    \label{eq:positives}
    \positives & = TP + FN \\
    \label{eq:negatives}
    \negatives & = \total - \positives \\
    \label{eq:prevalence_rate}
    \prevalence & = \frac{\positives}{\total} \\
    \label{eq:recall}TPR &=\frac{TP}{\positives}= \recall = \frac{TP}{TP + FN} \\
    \label{eq:false_alert_rate}FAR &=\frac{FP}{\negatives}= \frac{FP}{FP + TN} \\
    \precision & = \frac{TP}{TP + FP} \label{eq:precision}
\end{align}

$TP$, $FP$, $TN$, and $FN$, which are necessary to calculate the metrics of interest are respectively the True Positives, False Positives, True Negatives, and False Negatives. The TP represent the share of elements classified as positive which are positive while the FP represents the elements classified as positive which are negative. The TN are the elements classified as negative which are negative, and the FN are the elements classified as negative which are instead positive.

For our purposes, it is more useful to express $TP$, $FN$ and $FP$ in terms of the other values that are often found in publications reporting results of \aias\ components. 

\begin{proposition}
Let \recall\ be the recall of a classifier and  \precision\ be its precision. When applied to a domain with \total\ elements and a prevalence rate of \prevalence, the true positives $TP$, false negatives $FN$, and false positives $FP$ of the classifier are as follows:
\begin{align}
    \label{eq:TP:start} 
    TP & = \recall \cdot \prevalence \cdot \total \\
    \label{eq:FN:start} 
    FN &= (1-\recall) \cdot \prevalence \cdot \total\\
    \label{eq:FP:start}
    FP &= \recall \cdot \frac{1- \precision}{\precision} \cdot \prevalence \cdot \total 
\end{align}
\end{proposition}
\begin{proof}
\label{proof1}
    The first two equations are simply an inversion of the definition of recall (\ref{eq:recall}), where positives \positives\ are expressed as a function of the prevalence rate (\ref{eq:prevalence_rate}). The third equation is obtained by inverting the definition of precision (\ref{eq:precision}) to express false positives $FP$ as a function of $TP$ and \precision\ and then replace into it the equation computing $TP$ as a function of recall \recall\ and prevalence \prevalence (\ref{eq:TP:start}).
\end{proof}

\subsection{Deterministic recall, partial repairs, no breaking changes}
\label{sec:prederror:noperfectrepair:nobreak}
\begin{proposition}
\label{prop:prederror:noperfectrepair:nobreak} 
Let \recall\ be the recall rate of a classifier that is used both as a first and and second classifier, let \fixrate\ be the theoretical fix rate of the fixer which (i) only affects positive (vulnerable) code and (ii) does not break nor make vulnerable code of the not vulnerable code which is eventually piped through it. The classifier can also correctly recognize unsatisfactory fixes (iii) with the same \recall.  
Then the \aias\ true performance when applied to a domain with \total\ elements and an initial prevalence rate of \prevalence, is 
\begin{align}
    \fixaias & = \fixrate \cdot\recall \label{eq:fixaias} \\
    \prevalenceaias & = (1 - \fixrate\cdot\recall)\cdot \prevalence \\
    TPR(aias) &= \recall \cdot \frac{(1-\fixrate)\cdot \recall}{1 - \fixrate\cdot\recall } \\
    FAR(aias) & = \recall^2 \cdot \frac{1- \precision}{\precision}\frac{(1-\fixrate)\cdot\prevalence }{1 - (1 - \fixrate\cdot\recall)\cdot \prevalence}
\end{align}
\end{proposition}

Results show that, unless the fix rate is perfect, the final prevalence rate is not reduced to zero and it will depend on the uncertainty in the recall. 

An apparently surprising result is that if the fix rate is perfect then the overall true positive rate ($TPR$) is zero. This is actually to be expected: with a perfect fix rate, all identified positives are fixed. This does not mean that all positives are eliminated because the false negatives from the first classifier are still present. In general, since $\recall \leq 1$ we have that the term $\frac{(1-\fixrate)\cdot \recall}{1 - \fixrate\cdot\recall}\leq 1$ and therefore the recall of the \aias \@\xspace as a whole is lower than the recall of the first classifier, i.e.\ $TPR(aias)\leq TPR$ (see Appendix, section \ref{sec:tpr_aias_tpr}).

While the recall of the \aias \@\xspace does not depend on the prevalence rate, the false alert rate (FAR) depends in a non-linear way on the overall prevalence rate of the system. It is still possible to prove that the false alert rate of the \aias as a whole is lower than the false alert rate of the first classifier, i.e.\ $FAR(aias)\leq FAR$.  

\begin{proof}
\ref{prop:prederror:noperfectrepair:nobreak}
\label{proof2}
The first classifier receives in input the positives and the negatives and divides them into $TP$, $FP$, $FN$, and $TN$. 

The fixer receives in input $TP_{\nth{1}}+FP_{\nth{1}}$ of which only a fraction \fixrate\ of $TP_{\nth{1}}$ is actually fixed (Assumption (i)). 
According to assumption (ii) the fixer will not transform the false positive into new positives (i.e. it will not transform them into positives nor will not break them). Since the second instance of the classifier does not change the nature of the processed object but at worst misclassifies it we have that
\begin{align}
\label{eq:positive:end}
    \positives(aias) & = \overbrace{(1-\fixrate) TP_{\nth{1}}}^{\mbox{unfixed by fixer}} \hspace{4ex} + \hspace{-4ex} \overbrace{FN_{\nth{1}}}^{\mbox{misclassified by \nth{1} classifier}}  \\
    \label{eq:positive:aiases}
    \positives(aias) & = \positives - \fixaias \cdot \positives = \prevalence \cdot \total - \fixaias \cdot \prevalence \cdot \total
\end{align}
We now equate the terms, replace $TP_{\nth{1}}$ and $FN_{\nth{1}}$ with the corresponding equations, and simplify $\prevalence \cdot \total$ from both sides of the equation to obtain $1 - \fixaias =  (1-\fixrate) \cdot \recall + (1 - \recall)$ which simplifies to $\fixaias =  \fixrate \cdot \recall$ (see Appendix, section \ref{sec:pos_aias_derivation}).

We can use equation (\ref{eq:positive:aiases}) to directly obtain the prevalence rate for the \aias\ by replacing the value of \fixaias\ just computed and dividing by the total number of elements \total. 

To compute the true positive rate we replace in the definition of $TPR$ (\ref{eq:recall}) the number of $TP$ surviving at the end of the second classifier which is $(1-\fixrate)\cdot TP_{\nth{1}}\cdot \recall$ because by assumption (ii) and (iii) only the original true positives will be reclassified as positives. We divide by the total number of positives of the \aias\ as computed from equation \ref{eq:positive:aiases}. By simplifying both numerator and denominator for $\prevalence \cdot \total$ we obtain
\begin{align}
    TPR(aias) &= \frac{(1-\fixrate)\cdot \recall\cdot \recall}{1 - \fixrate\cdot \recall }
\end{align}

To compute the false alert rate we need to compute first the false positives of the second classifier. To this extent, we rewrite the definition of false positives (\ref{eq:FP:start}) in terms of the new set of positives $(1-\fixrate) TP_{\nth{1}}$ at the end of the fixer according to equation (\ref{eq:positive:end}).
\begin{align}
     FP(aias) &= \recall \cdot \frac{1- \precision}{\precision} \cdot (1-\fixrate)\cdot \recall \cdot \prevalence \cdot \total
\end{align}
Then we substitute this value into the definition of the false alert rate (\ref{eq:false_alert_rate}) with the value of the overall negatives of the system.
\end{proof}
\begin{corollary}
Unless the fix rate is perfect ($\fixrate=1$) the number of false negatives of the \aias\ satisfying the condition of Proposition \ref{prop:prederror:noperfectrepair:nobreak} is higher than the number of false negatives that would result from just the first classifier. The false negatives of the \aias\ also increase with the increase in recall \recall.
\end{corollary}
This result is surprising as we expected the system to improve as recall improves. However, a larger recall would also mean that more positives would be piped through the fixer and tested again. Since the fixer is not perfect the number of false negatives emerging from the second run of the classifier will increase.
\begin{proof} We compute the false negatives at the end of the \aias\ starting from the definition as 
\begin{align}
     FN(aias) &= \underbrace{\overbrace{\left[(1 - \fixrate)\cdot TP_{\nth{1}}\right]}^{\mbox{unfixed positives}}\cdot (1 - \recall)}_{\mbox{Escaping the \nth{2} classifier}} + FN_{\nth{1}} \label{eq:fnaias}
\end{align}
We plug in the definition of $TP_{\nth{1}}$ and $FN_{\nth{1}}$ in terms of positives \positives\ and thus of the prevalence rate \prevalence\ and the overall number of objects \total\ and re-arrange the terms to obtain
\begin{align}
     FN(aias) &= \left[1+(1 - \fixrate)\cdot \recall\right]\cdot (1 - \recall) \cdot \prevalence \cdot \total \\
              &= \left[1+(1 - \fixrate)\cdot \recall\right]\cdot FN_{\nth{1}} 
\end{align}
\end{proof}
By using the above equations we can compute the total number of elements which will be passed to the fixer and the second classifier.
\begin{align}
\label{eq:N:fixer}
     \total_{\nth{2}} &= TP_{\nth{1}} + FP_{\nth{1}} = \frac{\recall}{\precision} \cdot \prevalence \cdot \total
\end{align}
By using assumption (iii) the positive that will recognized as such $TP(aias)$ as 
\begin{align}
    TP_{\mbox{\nth{2}}} &= \underbrace{\overbrace{(1 - \fixrate)\cdot TP_{\nth{1}}}^{\mbox{unfixed positives}} \cdot \recall}_{\mbox{after the second classifier}}
\end{align}
By expanding the definition of $TP_{\nth{1}}=\recall\cdot \positives$ (\ref{eq:TP:start}) and the definition of positives as $\positives=\prevalence\cdot\total$ (\ref{eq:positives}) and re-arranging the terms we have
\begin{align}
    TP(aias) &= (1 - \fixrate) \cdot \recall^2 \cdot \prevalence \cdot \total
\end{align}

Then we can revise the final prevalence rate as
\begin{align}
    \prevalence(aias) &= \frac{TP(aias) + FN(aias)}{TP(aias) + FN(aias) + TN(aias) + FP(aias)} \\
    & = \frac{TP(aias) + FN(aias)}{\total} \nonumber
\end{align}
We can plug the solution for $TP(aias)$ and $FN(aias)$ and observe that
they are both multiplied by common factor $\prevalence\cdot\total$ which allows us to simplify the denominator and remove the dependency by the total number of objects. The ratio between the final prevalence and the initial prevalence rate is then captured by the following expression:
\begin{align}
\frac{\prevalence(aias)}{\prevalence} & =
(1 - \fixrate) \cdot \recall^2 + \left[1+(1 - \fixrate)\cdot \recall\right]\cdot (1 - \recall)
\end{align}
which further algebraically simplifies as follows:
\begin{align}
\prevalence(aias) & = \left(1 - \fixrate\cdot \recall\right)\cdot \prevalence \label{eq:prevalence_aias} 
\end{align}
By multiplying both ends by \total\ we obtain the total number of positives before and after the treatment by the \aias\ pipeline. The \aias fix rate is therefore equal to
\begin{align}
\frac{\positives - \positives(aias)}{\positives} & = \fixrate\cdot \recall
\end{align}

Complete derivations of $\prevalenceaias$, $TPR(aias)$, and $FAR(aias)$, $\total_{\nth{2}}$, $FN(aias)$, $TP(aias)$, in sections \ref{sec:pr_end_derivation}, \ref{sec:tpr_aias_derivation}, \ref{sec:far_aias_derivation}, \ref{sec:n2_derivation}, \ref{sec:fn_derivation}, \ref{sec:tp_derivation}, in the Appendix.

\subsection{Uncertain recall}
\label{sec:uncertainrecall}
 
Considering the low availability of recall values, we do not have enough data to approximate a specific cumulative distribution function (CDF) (e.g. beta, normal, etc.) (Ferson et al., \citeyear{Ferson2013129}; Gray et al., \citeyear{gray2019problem}). We thus rely on distribution-free analysis (Gray et al., \citeyear{gray2022distribution}). Specifically, to model uncertainty in the recall and propagate it in the form of intervals, we employ nonparametric probability boxes (p-boxes) and probabilistic bound analysis (PBA) (Ferson et al., \citeyear{ferson1996different}, \citeyear{ferson2003constructing}; Iskandar \citeyear{iskandar2021probability}).

By substituting a specific CDF with p-boxes, PBA allows to model the lack of knowledge regarding the specific CDF from which the recall values are sampled. Considering that we cannot possess exhaustive information regarding the CDF of recalls of AI vulnerability detectors, the choice of this mathematical tool is preferred, compared to the use of precise probability density functions. Thus, we employ non-parametric p-boxes which allows to model uncertainty when the shape of the distribution is not known but the parameters of the CDF are known such as the $min$, $max$, and $\mu$, which respectively correspond to the minimum, maximum, and expected value of the random variable.

Equations (\ref{form:p-box1}) and (\ref{form:p_box2}), are the inverse p-boxes derived by Iskandar (\citeyear{iskandar2021probability}), that substitute the inverse of the specific CDF and thus are used to sample the lower and upper bound recall values.

\begin{align}
    \label{form:p-box1}
    \underline{rec}(p)^{-1}_{a, b, \mu} = 
    \begin{cases}
        [a, \mu] \ \quad for \ p = 0 \\
        \frac{p \cdot a - \mu}{p - 1} \ \ for \ 0 < p < \frac{b - \mu}{b - a} \\
        b \ \ \quad \quad for \ \frac{b - \mu}{b - a} \leq p \leq 1 \\
    \end{cases}
\end{align}

\begin{align}
    \label{form:p_box2}
    \overline{rec}(p)^{-1}_{a, b, \mu} = 
    \begin{cases}
        a \ \ \quad \quad \quad for \ 0 \leq p \leq \frac{b - \mu}{b - a} \\
        b - \frac{b - \mu}{p} \ \ for \ \frac{b - \mu}{b - a} < p < 1 \\
        [\mu, b] \ \ \quad for \ p = 1 \\
    \end{cases}
\end{align}

where $\underline{rec}$ stands for the inverse of the p-box that models the recall, and $a$, $b$, and $\mu$, respectively correspond to the minimum, the maximum and expected value of the recall registered during the literature review (Section \ref{sec:recall_in_the_field}). $p$ is the value sampled from the standard uniform distribution and given in input to the inverse probability box to sample the recall value. Sampling recall values from the lower and upper bounds of the inverse of the probability boxes allows treating recall as an interval, consisting of a minimum and a maximum possible values. Now, employing an intervalised recall will have as a consequence the formation of intervals in all the equations in which the recall is used. Respectively, the use of intervalised recall in Equation (\ref{eq:prevalence_aias}), (\ref{eq:fixaias}), and (\ref{eq:fnaias}), will lead to the generation of intervalised final prevalence rate, final fix rate and final false negatives ratio.

\section{Recall in the field}
\label{sec:recall_in_the_field}
We collect the reported recall values (and precision) of AI-augmented vulnerability detectors and derive the parameters necessary to implement the p-boxes in our simulations. 

\subsection{Search in digital libraries}
\label{sec:search_in_digital_libraries}
\begin{figure*}[h]
    \centering
    \includegraphics[width=\textwidth]{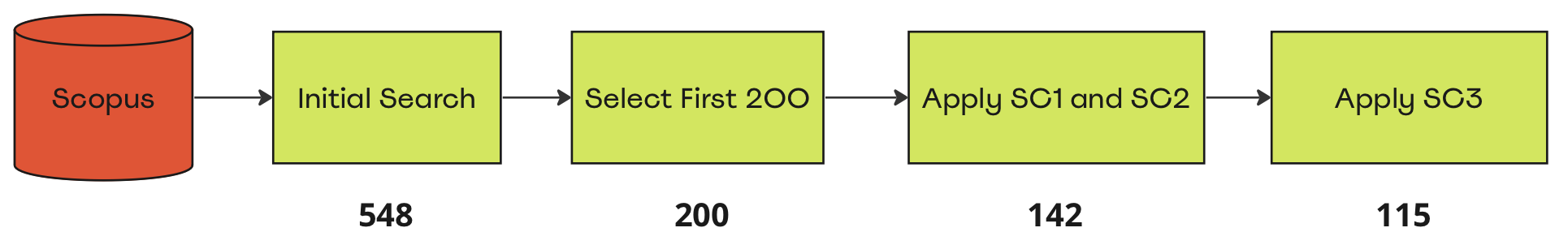}
    \caption{The figure shows the steps implemented to gather the publications from which to extract recall values. The first step consists of an initial search on Scopus that retrieves 548 publications. Based on Scopus relevance's ranking, the first 200 are selected (each 50 down the ranking, the fraction of relevant papers drops significantly and no relevant paper was found from 200 to 250). We check which of these 200 publications implement vulnerability detection or fault localization using AI (SC1 and SC2). In the end, of the 142 publications that respect the previous conditions, we select only the ones which used recall as an evaluation metric (SC3).}
    \label{fig:review_steps}
\end{figure*}

Figure \ref{fig:review_steps} illustrates the steps that define our search. We defined a search string to filter publications stored in digital libraries: \textbf{(``vulnerability detection'' OR ``fault localization'') AND (``artificial intelligence'' OR ``AI'' OR ``deep learning'' OR ``DL'' OR ``machine learning'' OR ``ML'') AND (``sensitivity'' OR ``true positive rate'' OR ``TPR'' OR ``recall'' OR ``hit rate'') AND ``code''}.

We define a list of selection criteria (SC) that a publication must respect to be selected for the extraction of data points.

\begin{itemize}
    \item SC1. The publication must be related to the topic of vulnerability detection or fault localization. For instance, we discard publications related to general \emph{feature location}.
    
    \item SC2. The publication must apply ML or DL algorithms to the problem of vulnerability detection. We discard the publications which do not employ ML or DL.
    
    \item SC3. Since the metrics considered $\prevalenceaias$, $\fixaias$, and $FN(aias)$, depend uniquely on the recall, the publication must (at least) report the recall of the vulnerability detectors.
\end{itemize} 

By employing the search string on Scopus, the resulting total number of publications is 548. The following selection of the papers is guided by the standards of the Preferred Reporting Items for Systematic Review and Meta-Analysis (PRISMA) Statement (Page et al., \citeyear{page2021prisma}), which suggests relevance as a selection method. Therefore we use the built-in relevance score provided by Scopus which ranks publications based on their affinity with the presented search string (Elsevier, \citeyear{relevanceDefinition}). We empirically found that after the \nth{200} publication, the selected publications do not either concern the problem of vulnerability detection or the application of AI, Deep Learning or Machine Learning to the problem. By looking at the \nth{1} paper to the \nth{50} we selected 45 papers respecting all success criteria (90\% of the scanned sample), from the \nth{51} to the \nth{100} we selected 35 papers matching the criteria (70\%), from the \nth{101} to the \nth{150} 25 papers (50\%), and from the \nth{151} to the \nth{200} 11 papers (22\%). We kept analyzing until the \nth{250} paper and found no publication that respected the selected criteria. Therefore we stopped the search and considered the first 200 papers. After applying SC1 and SC2 to the title and abstract, we retain 142 publications. Finally, applying SC3 resulted in removing 26 more publications.

\subsection{Collected samples}
\label{sec:collected_samples}
\label{subsubsec:explor_data_analysis}
For each article, we select the recall value of the model presented in the publication.
We also select the recall values related to baseline models, but only if those values are derived from new experiments. If the values are simply reported from the publications where baseline models are presented, we consider them duplicates. We include recall values related to the same model used in different publications because as a consequence of repeated experiments, the model performance can differ between different studies. The factors responsible for different performances for the same model are the following:  

\begin{itemize}
    \item Different dataset: the dataset used by the new paper on which the new model is tested and compared to old models can be different compared to the dataset on which previous models were originally tested. 
    \item Different training modalities: if the authors of the new paper retrain all the models and change the training modalities, this will impact the models' performance. 
    \item Random changes: even when adopting the same training techniques other factors can influence the final training result, such as the random training-test splitting and the hardware on which experiments are performed. 
\end{itemize}

From an initial sample of 2328 values, eliminating the values not derived by new experiments and the outliers, we obtain 2227 samples that we use to calculate the p-boxes parameters for the simulation. We eliminate outliers by employing z-scores. Specifically, if the recall data point possesses a z-score greater or equal to 3 or a z-score smaller or equal to -3 we consider it to be an outlier (Chen et al., \citeyear{chen2022combining}). The minimum and maximum reported recall are respectively 0.06 and 1.00, while the mean is 0.75. For completeness in Table \ref{tab:descriptive} we also show descriptive statistics on the collected precision samples (but we do not use them yet in our simulation).

\begin{table*}[h]
\caption{Descriptive statistics regarding recall and precision data, gathered from publications related to the applications of AI to vulnerability detection. For both recall and precision, the table reports the minimum and maximum value registered, the first quartile (Q1) the third quartile (Q3), the mean, the median and the standard deviation (SD).}
\label{tab:descriptive}
\centering
\begin{tabular}{llllllllll}
\hline
\textbf{Measure} & \textbf{Samples} & \textbf{Selected} & \textbf{Min} & \textbf{Q1} & \textbf{Median} & \textbf{Q3} & \textbf{Max} & \textbf{Mean} & \textbf{SD} \\ \hline
Recall & 2227 & 116 & 0.06 & 0.62 & 0.80 & 0.92 & 1.00 & 0.75 & 0.21 \\
Precision & 2016 & 100 & 0.00 & 0.56 & 0.78 & 0.92 & 1.00 & 0.71 & 0.27 \\ \hline
\end{tabular}
\end{table*}

Regarding the True Negative Rate $(TNR)$ and the False Positive Rate $(FPR)$, where $TNR = 1 – FPR$, among the selected publications only two report the $TNR$ and only 14 publications report the $FPR$. This remains a significant limitation of the data reported in the literature, so it is difficult to understand the trade-off faced by the studies.

\section{Simulation One: Constant prevalence rate}
\label{sec:simulation}
Through the simulation, we are interested in calculating $\prevalenceaias$, $\fixaias$, $FN(aias)$ which, as previously shown (Section \ref{sec:methodology}), depend uniquely on the recall. To allow for future extensions, we implemented the simulator taking into account $TN$ and $FP$, which are needed to define specificity. At this stage of the research, the specificity value does not affect the final result, thus we set its value to zero. We implement the simulation through the \emph{pba-for-python} library as it allows to perform rigorous p-box arithmetic (Gray et al., \citeyear{gray2022probability}). Additionally, to show the influence that the number of samples has on the precision of the simulation, we implement the experiments also through Monte Carlo (MC) simulation (Metropolis \& Ulam, \citeyear{metropolis1949monte}) and report the results in the Appendix in Sections \ref{sec:appendix:pr_aias} \ref{sec:appendix:fr_aias}, \ref{sec:appendix:fn_aias}, \ref{sec:appendix:second:fr_aias:constant}, and \ref{sec:appendix:second:fr_aias:no_constant}.
 
\subsection{Simulator}
\begin{figure*}[h]
    \centering
    \includegraphics[width=\textwidth]{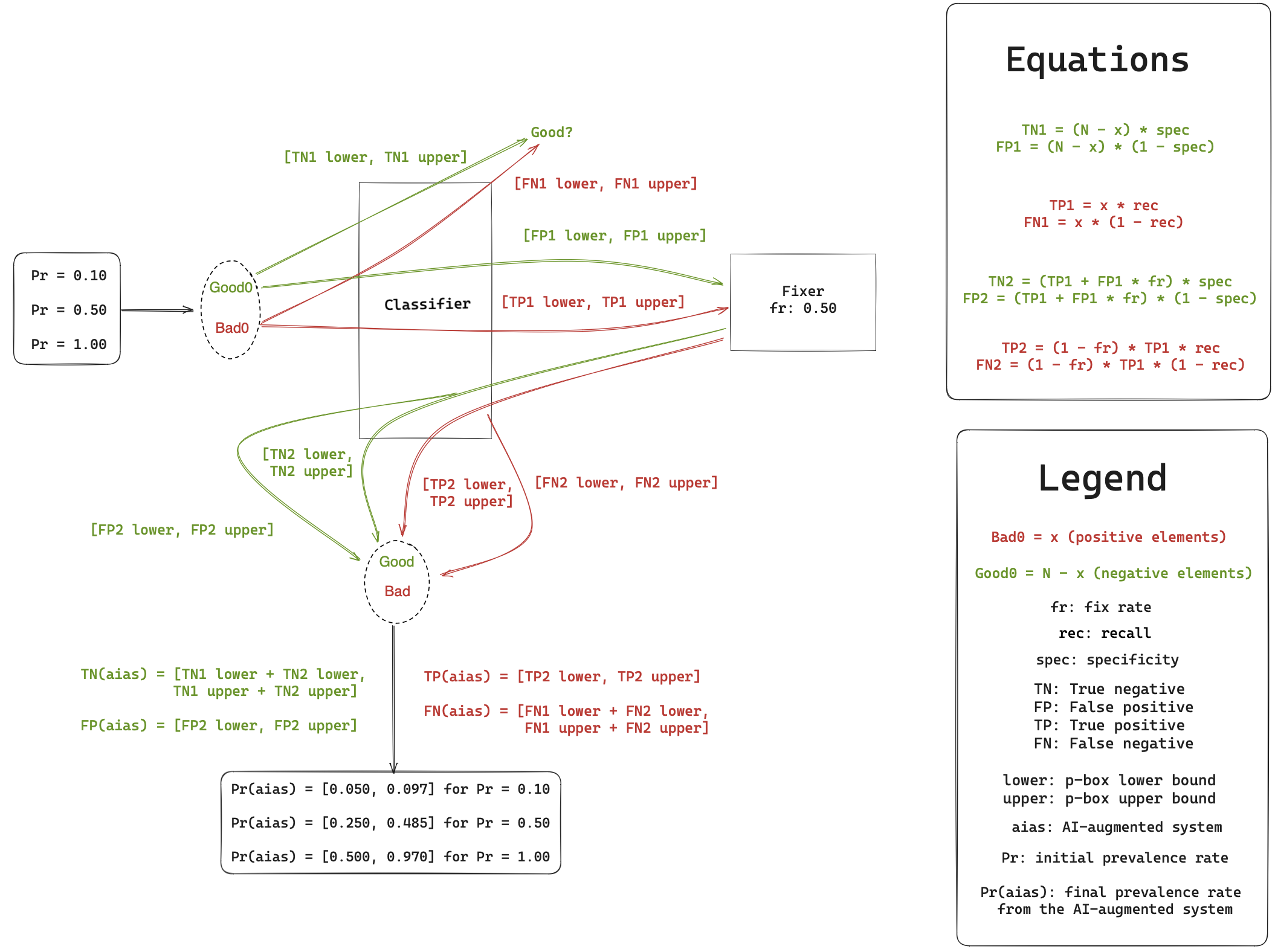}
    \caption{Illustration of the process that leads to the calculation of the final prevalence rate ($\mathbf{\prevalenceaias}$), given a fixed rate of $0.50$ and three different starting prevalence rates ($\mathbf{\prevalence}$). $\prevalence$ determines the number of positives in the ground truth, while $\prevalenceaias$ is the ratio between the positives ($\mathbf{TP(aias) + FN(aias)}$) and the total elements ($\mathbf{N}$) at the end of the process. We represent the pipeline as a loop because the first and the second classifiers possess the same recall and specificity. By considering at each step in the pipeline, a lower and upper bound of the recall, we propagate the uncertainty, with the consequence that also the $\prevalenceaias$, as $TP$, $FN$, $TN$, and $FP$, will have an upper and lower bound. The lower bound of the $\prevalenceaias$ is the best-case scenario, which is the case in which the classifier is perfect. In reality, $\prevalenceaias$ can be equal to all the values contained in the interval, depending on the classifier's performance.}
    \label{fig:pr}
\end{figure*}

Figure \ref{fig:pr} illustrates the subsystems of our simulation pipeline. It comprises a fixer and a classifier that acts as the first and second classifiers.

\definecolor{codegreen}{rgb}{0,0.6,0}
\definecolor{codegray}{rgb}{0.5,0.5,0.5}
\definecolor{codepurple}{rgb}{0.58,0,0.82}
\definecolor{backcolour}{rgb}{0.95,0.95,0.92}
\definecolor{backcolour}{rgb}{1.00,1.00,1.00}

\lstdefinestyle{mystyle}{
    deletekeywords=[2]{not, pass},
    backgroundcolor=\color{backcolour},   
    commentstyle=\color{codegreen},
    keywordstyle=\color{magenta},
    numberstyle=\tiny\color{codegray},
    stringstyle=\color{codepurple},
    basicstyle=\ttfamily\footnotesize,
    breakatwhitespace=false,         
    breaklines=true,                 
    captionpos=b,                    
    keepspaces=true,                 
    numbers=left,                    
    numbersep=5pt,                  
    showspaces=false,                
    showstringspaces=false,
    showtabs=false,                  
    tabsize=2
}

\lstset{style=mystyle}

\subsubsection{Ground truth generator} 
\label{sec:groun_truth_gen}
The ground truth generator creates the dataset that allows the simulation of the pipeline. Each generated element represents a code sample, which can be vulnerable or not vulnerable. Thus, the ground truth generator produces fictional positive and negative elements ($Pos$, $Neg$): 

\begin{itemize}
    \item It receives as input the total number of elements ($N_{E}$), set to $100,000$, and the initial prevalence rate ($\prevalence$), which defines the initial number of vulnerable elements.
    
    \item The generator labels each object as vulnerable with probability equal to $\prevalence$, and not vulnerable with probability $1 - \prevalence$ and returns a list containing all the samples generated. 
\end{itemize}

\subsubsection{P-boxes and recall sampling}
\label{sec:p-boxes-and-recall-sampling}

We employ the \emph{pba-for-python} library to sample $N_{R}$ lower and $N_{R}$ upper bound recall values, where $N_{R} = 202$ (default value set by the \emph{pba-for-python} library):

\begin{itemize}
    \item The parameters used to sample from p-boxes formulas as the minimum $(0.06)$, maximum $(1.00)$, and mean $(0.75)$ value generated from the exploratory data analysis in Section \ref{subsubsec:explor_data_analysis}. 

    \item Given two lists of recall values, one representing the lower bounds and the other the upper bounds, each of $N_{R}$ samples, we perform the simulation to estimate the upper and lower bounds of the metrics of interest.
\end{itemize}

\subsubsection{First classifier} 
After generating the lower and upper bound recall values, the first classifier executes the first subdivision of the samples, generating $TP_{\nth{1}}$, $FN_{\nth{1}}$, $TN_{\nth{1}}$, and $FP_{\nth{1}}$. 

\begin{itemize}
    \item The first classifier discerns each vulnerable element of the ground truth between $TP$ with probability equal to \recall\ and as $FN$ with probability equal to $1 - \recall$. This means that the greater the recall the greater the probability that vulnerable objects are classified as $TP$.
    
    \item Since the first classifier is simulated with both lower and upper bound recall values, in the end, we obtain lower and upper bounds for each element, thus $[\underline{TP_{\nth{1}}}, \overline{TP_{\nth{1}}}]$, $[\underline{FN_{\nth{1}}}, \overline{FN_{\nth{1}}}]$, $[\underline{TN_{\nth{1}}}, \overline{TN_{\nth{1}}}]$, $[\underline{FP_{\nth{1}}}, \overline{FP_{\nth{1}}}]$. 
\end{itemize}

\subsubsection{Fixer} 
The fixer, with fix rate $\fixrate$, tries to repair the samples classified as positives by the first classifier, namely $TP_{\nth{1}}$ and $FP_{\nth{1}}$. The fixer repairs each sample classified as positive with probability equal to $\fixrate$. Since we assume that a $FP$ cannot be broken, the intervention on $FP$ cannot cause it to become a $TP$.

\subsubsection{Second classifier} 
The second classifier, with the same recall and specificity as the first classifier, classifies the objects that passed through the fixer, generating $TP_{\nth{2}}$, $FN_{\nth{2}}$, $TN_{\nth{2}}$, and $FP_{\nth{2}}$. 

\begin{itemize}
    \item The second classifier labels each vulnerable object that passed through the fixer as $TP$ with probability equal to the \recall\ and as $FN$ with probability $1 - \recall$.
    
    \item Since the second classifier is simulated with both lower and upper bound recall values, we obtain lower and upper bounds for each element, thus $[\underline{TP_{\nth{2}}}, \overline{TP_{\nth{2}}}]$, $[\underline{FN_{\nth{2}}}, \overline{FN_{\nth{2}}}]$, $[\underline{TN_{\nth{2}}}, \overline{TN_{\nth{2}}}]$, and $[\underline{FP_{\nth{2}}}, \overline{FP_{\nth{2}}}]$
\end{itemize}

\subsubsection{Final counter} 
The final counter gathers the results from the first classifier, the fixer, and the second classifier and that calculates the final prevalence rate ($\prevalenceaias$), the final fix rate ($\fixaias$), and the false negatives ratio ($FN_{ratio}$) between the final number of false negatives ($FN(aias)$) and the false negatives generated by the first classifier ($FN_{\nth{1}}$). Since the uncertainty propagates until the final counter, each metric will be characterized by a lower and upper bound, thus: $[\underline{\prevalenceaias}, \overline{\prevalenceaias}]$, $[\underline{\fixaias}, \overline{\fixaias}]$, $[\underline{FN_{ratio}}, \overline{FN_{ratio}}]$. 

\subsection{Simulation results}
We present the simulation results and show how propagating uncertainty affects $\prevalenceaias$ (see Table \ref{tab:pr_end}), $\fixaias$ (see Table \ref{tab:fr}), and the false negatives (see Table \ref{tab:fn}). Figure \ref{fig:pr} instantiates the simulated pipeline, with the results obtained from the simulation with $\prevalence = 0.50$ and $\fixrate = 0.50$.

\subsubsection{Final prevalence rate} 
\label{sec:final_prev_rate}
Table \ref{tab:pr_end} shows the results related to the decrease in the prevalence rate. We run simulations with $\prevalence = (0.10, 0.50, 1.00)$, thus in the first, second and third sets of simulation, the total number of vulnerable samples is equal to the $10\%$, $50\%$ and $100\%$ of the total samples. For each of these simulation sets, we calculate the final prevalence rate with $\fixrate = (0.50, 0.70, 0.90, 1.00)$, meaning that the expected decrease in the prevalence rate is respectively $50\%$, $70\%$, $90\%$ and $100\%$. But, the theoretical decrease in the prevalence rate that should be observed given a specific starting prevalence rate and fix rate, is only the lower bound of the interval, which corresponds to the minimum prevalence rate obtainable when the capacity to locate vulnerable elements is perfect. In all the other cases the value will fall within the bounds of the interval. For example, when $\prevalence = 0.50$ and $\fixrate = 0.50$ we should observe a decrease in the final prevalence rate of $50\%$, thus $\prevalenceaias = 0.250$. But, Table \ref{tab:pr_end} and Figure \ref{fig:pr} show that the $50\%$ decrease only represents the lower bound, contrasting with an upper bound of $0.485$.

\begin{table}[h]
\centering
\caption{This table shows how the bounds of the final prevalence rate ($\mathbf{\prevalenceaias}$) given initial prevalence rate ($\mathbf{\prevalence}$), and theoretical fix rate ($\mathbf{\fixrate}$), but uncertain recall. The theoretical decrease of the initial prevalence rate given a fix rate only consists of the lower bound of the interval, which is when recall is equal to one. For instance when $\mathbf{\prevalence = 1.00}$ and $\mathbf{\fixrate = 0.50}$, the prevalence rate decreases of the $0.50\%$ but only as a lower bound. Section \ref{sec:appendix:pr_aias} in the Appendix, presents the results of the calculation of the final prevalence rate obtained through MC simulation.} 
\label{tab:pr_end}
\begin{tabular}{@{}lllll@{}}
\hline
\multicolumn{1}{c}{\textbf{}}    & \multicolumn{4}{c}{$\mathbf{\prevalenceaias}$}                                                      \\ \hline
\multicolumn{1}{l|}{$\mathbf{\prevalence}$} & $\mathbf{\fixrate = 0.50}$ & $\mathbf{\fixrate = 0.70}$ & $\mathbf{\fixrate = 0.90}$ & $\mathbf{\fixrate = 1.00}$ \\ \hline
\multicolumn{1}{l|}{0.10}        & {[}0.050, 0.097{]}   & {[}0.030, 0.096{]}   & {[}0.010, 0.095{]}   & {[}0.000, 0.094{]}   \\
\multicolumn{1}{l|}{0.50}        & {[}0.250, 0.485{]}   & {[}0.150, 0.479{]}   & {[}0.050, 0.473{]}   & {[}0.000, 0.470{]}   \\
\multicolumn{1}{l|}{1.00}        & {[}0.500, 0.970{]}   & {[}0.300, 0.958{]}   & {[}0.100, 0.946{]}   & {[}0.000, 0.940{]}   \\ \hline
\end{tabular}
\end{table} 

\subsubsection{Real fix rate} Table \ref{tab:fr} show the results related to the final fix rate. We run simulations with theoretical fix rate $\fixrate = (0.50, 0.70, 0.90, 1.00)$. At the end of the simulations, $\fixrate$ only corresponds to the upper bound of the interval of $\fixaias$, which is the maximum fix rate obtainable when the capacity to locate vulnerable elements is maximum. For example, when $\fixrate = 0.50$, the $\fixaias$ oscillates between a maximum of $0.500$ equal to $\fixrate$ and a minimum of $0.030$. This illustrates the limitations of APR tools and the importance of stating the final results in terms of intervals and not of single numbers in order to represent the uncertainty that characterizes these systems when they are applied to real-world scenarios. 

\begin{table}[h]
\caption{Comparison between the theoretical fix rate ($\mathbf{\fixrate}$) and the real fix rate ($\mathbf{\fixaias}$), when $\mathbf{\prevalence = 0.50}$. The theoretical fix rate only translates into the upper bound of the interval, while the real fix rate can fall within a much wider range of values, which will eventually depend on the quality of the classifier. Section \ref{sec:appendix:fr_aias} in the Appendix presents the resulting final fix rate obtained through MC simulation.}
\centering
\label{tab:fr}
\begin{tabular}{ll}
\hline
$\mathbf{\fixrate}$ & $\mathbf{\fixaias}$ \\ \hline
0.50 & {[}0.030, 0.500{]} \\
0.70 & {[}0.042, 0.700{]} \\
0.90 & {[}0.054, 0.900{]} \\
1.00 & {[}0.060, 1.000{]} \\ \hline
\end{tabular}
\end{table}

\subsubsection{False negatives ratio} 
Table \ref{tab:fn} shows the results related to $FN_{ratio}$, which is the ratio between the false negatives generated by the first classifier $FN_{\nth{1}}$ and the overall number of false negatives registered at the end of the pipeline $FN(aias)$. Apart from $\fixrate = 1$, the final ratio is always greater than one, and this indicates that the pipeline is unable to avoid the growth of the number of $FN$ between the first and the second classifier. 

\begin{table}[h]
\caption{This table shows the final bounds regarding the ratio ($\mathbf{FN_{ratio}}$). Between the first and the second classifier, the number of $FN$ grows apart in the case in which the $\mathbf{\fixrate = 1}$. When the $\mathbf{\fixrate = 1}$, $\mathbf{FN(aias) = FN_{\nth{1}}}$, because there will be no positives that can be classified as $FN$ by the second classifier and thus the number will not increase, leaving the ratio equal to one. Section \ref{sec:appendix:fn_aias} in the Appendix presents the results related to the $FN_{ratio}$ obtained through MC simulation.}
\centering
\label{tab:fn}
\begin{tabular}{ll}
\hline
$\mathbf{\fixrate}$ & $\mathbf{FN_{ratio}}$ \\ \hline
0.50 & {[}1.000, 1.030{]} \\
0.70 & {[}1.000, 1.018{]} \\
0.90 & {[}1.000, 1.006{]} \\
1.00 & {[}1.000, 1.000{]} \\ \hline
\end{tabular}
\end{table}

\section{Simulation Two: Beyond constant prevalence rate}
\label{sec:simulation_two}
In the general case, the constant prevalence rate assumption, underlying the first simulation, does not hold. It is possible to get an expected number of $TP$, but impossible to know which positives are actually $TP$, when the classifier is applied to actual data. Only by applying the classifier on the field it is possible to know whether the positives are $TP$ or $FP$. The data of the simulation can be used to train the classifier and calculate its recall, which would be a characteristic (fixed value) of the classifier. If the classifier is applied to a different dataset, it is incorrect to just calculate $TP$ from the definition of recall. 

In this simulation, we aim to analyze the effects of removing the constant prevalence rate assumption on the final fix rate of the pipeline. To relax the assumption, we calculate $TP$, $FN$, $TN$ and $FP$, relying on the notion of Positive Predicted Value (PPV) and Negative Predicted Value (NPV) (Parikh et al., \citeyear{parikh2008understanding}; Gray et al., \citeyear{gray2020no}). PPV and NPV are defined as follows: 

\begin{align}
    PPV &= \frac{\recall \cdot \prevalence}{\recall \cdot \prevalence + (1 - \specificity) \cdot (1 - \prevalence)} \\
    NPV &= \frac{\specificity \cdot (1 - \prevalence)}{\specificity \cdot (1 - \prevalence) + (1 - \recall) \cdot \prevalence}
\end{align}

where $\recall$ corresponds to the $recall$ or $sensitivity$ of the classifier, $\prevalence$ is the prevalence rate of the dataset and $\specificity$ is the specificity of the classifier. Then, we use $PPV$ to calculate the number of $TP$ and $FP$ and $NPV$ to calculate $TN$ and $FN$ as follows: 

\begin{align}
    TP &= PPV \cdot \positives \\
    FP &= (1 - PPV) \cdot \positives \\
    TN &= NPV \cdot \negatives \\
    FN &= (1 - NPV) \cdot \negatives
\end{align}

where $\positives$ are the elements classified as positive and $\negatives$ are the elements classified as negatives. 

Differently from the first simulation, we assume the $min$, $max$, and $\mu$ parameters for the p-boxes. This allows us to employ p-boxes to sample $specificity$ values, as using assumed parameters removes the limitation posed by the lack of $specificity$ values reported in the literature. We measure the performance of the simulated APR tool, with three different $min$ values for $sensitivity$ and $specificity$ $0.50$, $0.70$, and $0.90$, and the $max$ value of $1.00$, measuring how raising the minimum value of recall and specificity will impact the final $\fixrate$ of the pipeline. We chose those values because they allow to cover the recall range from the first quartile to the third quartile (Table \ref{tab:descriptive}), including also the case of perfect recall. For specificity, we use the same values because we do not have enough values to make an informed choice. 

\subsection{Results of the simulation}
\label{sec:simulation_two_results}
\begin{table}[h]
\centering
\caption{This table shows for each theoretical fix rate ($\mathbf{\fixrate}$) how increasing the minimum recall and specificity of the classifier, affects the lower bound of the final fix rate ($\mathbf{\fixaias}$), in the case in which we maintain the assumption related to the consistency of the prevalence rate between training and test datasets. Section \ref{sec:appendix:second:fr_aias:constant} in the Appendix shows the results related to the $\fixaias$, obtained through MC simulation and with constant prevalence rate.}
\label{tab:const_prev_analysis_2}
\begin{tabular}{@{}llll@{}}
\hline
 & \multicolumn{3}{c}{\textbf{$\mathbf{\fixaias}$}} \\ \hline
\multicolumn{1}{c}{\textbf{}} & \multicolumn{3}{c}{\textbf{Min. Rec and Spec}} \\ \hline
\multicolumn{1}{c|}{\textbf{$\mathbf{\fixrate}$}} & \textbf{Min = 0.50} & \textbf{Min = 0.70} & \textbf{Min = 0.90} \\ \hline
\multicolumn{1}{l|}{0.50} & {[}0.250, 0.500{]} & {[}0.350, 0.500{]} & {[}0.450, 0.500{]} \\
\multicolumn{1}{l|}{0.70} & {[}0.350, 0.700{]} & {[}0.490, 0.700{]} & {[}0.630, 0.700{]} \\
\multicolumn{1}{l|}{0.90} & {[}0.450, 0.900{]} & {[}0.630, 0.900{]} & {[}0.810, 0.900{]} \\
\multicolumn{1}{l|}{1.00} & {[}0.500, 1.000{]} & {[}0.700, 1.000{]} & {[}0.900, 1.000{]} \\ \hline
\end{tabular}
\end{table}

\begin{table}[h]
\centering
\caption{This table shows for each theoretical fix rate $\mathbf{\fixrate}$ how increasing the minimum recall and specificity of the classifier, affects the lower bound of the final fix rate ($\mathbf{\fixaias}$), in the case in which we relax the assumption related to the consistency of the prevalence rate between the training and test datasets. Section \ref{sec:appendix:second:fr_aias:no_constant} in the Appendix presents the results related to the $\fixaias$ obtained through MC simulation when relaxing the assumption regarding the constant prevalence rate.}
\label{tab:non_cost_prev_analysis_2}
\begin{tabular}{@{}llll@{}}
\hline
 & \multicolumn{3}{c}{$\mathbf{\fixaias}$} \\ \hline
\multicolumn{1}{c}{\textbf{}} & \multicolumn{3}{c}{\textbf{Min. Rec and Spec}} \\ \hline
\multicolumn{1}{c|}{\textbf{$\mathbf{\fixrate}$}} & \textbf{min = 0.50} & \textbf{min = 0.70} & \textbf{min = 0.90} \\ \hline
\multicolumn{1}{l|}{0.50} & {[}0.000, 0.500{]} & {[}0.331, 0.500{]} & {[}0.500, 0.614{]} \\
\multicolumn{1}{l|}{0.70} & {[}0.000, 0.700{]} & {[}0.334, 0.700{]} & {[}0.684, 0.700{]} \\
\multicolumn{1}{l|}{0.90} & {[}0.000, 0.754{]} & {[}0.446, 0.787{]} & {[}0.779, 0.874{]} \\
\multicolumn{1}{l|}{1.00} & {[}0.000, 0.756{]} & {[}0.538, 0.794{]} & {[}0.877, 0.911{]} \\ \hline
\end{tabular}
\end{table}

Table \ref{tab:const_prev_analysis_2} and Table \ref{tab:non_cost_prev_analysis_2} respectively show the resulting fix rate, obtained by maintaining a constant prevalence rate and by relaxing the assumption. 

The results show that accounting for the shift in the prevalence rate modifies the final estimates of the $\fixaias$, downgrading what we can expect from the overall pipeline performance. For instance, examining the case in which the $\fixrate = 0.90$, comparing the results obtained considering the constant prevalence and shifted prevalence rate, the lower bound of the resulting $\fixaias$ is always higher when the prevalence rate is constant: when minimum recall and specificity are $0.50$, the lower bound for constant prevalence rate is $0.450$ and is lowered to $0.000$ when accounting for non-constant prevalence rate when recall and specificity are $0.70$, the lower bounds are respectively $0.630$ and $0.450$, and when minimum recall and specificity are 0.90 the lower bounds are respectively $0.810$ and $0.779$. 

This points to the necessity to account for possible variation in the prevalence rate of the dataset, by calculating the $TP$, $FP$, $TN$, and $FN$ through which $\fixaias$ is calculated employing the $PPV$ and $NPV$, to get a more realistic estimate of the capacities of the pipeline. 

We also see how progressively raising the minimum recall and specificity affects the final lower bound of the fix rate, both in the case of a constant prevalence rate and in the case in which the assumption is removed. For example, when the theoretical fix rate is $0.70$ and the minimum recall and specificity are $0.50$, the lower bound of the final fix rate is $0.350$, while raising the minimum value of recall and specificity to $0.70$ and then $0.90$, makes the lower bound grow to $0.490$ first and then $0.630$. The same can be said when the prevalence rate is not consistent and the $TP$, $FP$, $TN$, and $FN$ are calculated by employing the $PPV$ and $NPV$. When the theoretical fix rate is $0.70$, the final lower bound of the final fix rate increases from $0.000$, when the minimum recall and specificity are $0.50$, to $0.684$, when the minimum recall and specificity are $0.90$.

\section{Case study: AI-based APR}
\label{sec:case_study2}

We present a case study to measure the impact of uncertainty on AI-based APR tools.

This case study examines the possibility of obtaining an AI-augmented APR tool, composed of two AI subsystems, one dedicated to vulnerability detection, and the other to vulnerability repair. 

We analyze a DL-based APR tool, AIBUGHUNTER (Fu et al., \citeyear{fu2024aibughunter}). This pipeline is the result of the assembly of two systems, namely LineVul (Fu \& Tantithamthavorn, \citeyear{fu2022linevul}), which performs vulnerability detection and VulRepair (Fu et al., \citeyear{fu2022vulrepair}), which performs bug-fixing. Since the authors specified that they did not evaluate the whole AIBUGHUNTER pipeline in the dedicated publication, but that they evaluated the two composing tools separately, we use this case study to show to what extent uncertainty can impact the overall performance of an APR pipeline composed by different AI subsystems, trained on different datasets. We consider the dataset on which AIBUGHUNTER is tested, composed of $879$ total code samples, all of which have vulnerabilities. We calculate the number of the samples that the first classifier of the pipeline highlights to be vulnerable by multiplying the total code samples by the recall reported in the publication dedicated to LineVul (Fu \& Tantithamthavorn, \citeyear{fu2022linevul}) which amounts to $0.86$, obtaining $756$ \emph{Bad} code samples. Then, VulRepair (Fu et al., \citeyear{fu2022vulrepair}), with a reported repairing accuracy of $0.44$ is used to correct the bugs. Thus we multiply the repairing accuracy by the number of \emph{Bad} code samples, obtaining $333$ Fixed code samples. Thus the number of positive elements which the pipeline does not correct is equal to $423$. We then use our simulation pipeline to account for uncertainty in the recall, considering the same number of code samples and the same point estimate for repairing accuracy. When accounting for uncertainty the final repairing accuracy can be as high as $0.470$, and as low as $0.030$, compared to the starting point estimate of $0.44$.

\section{Discussion}

\subsection{Summary of results}
\label{sec:summary_results}
The results of the first simulation show that, once the uncertainty in the recall of the vulnerability detectors is propagated through the pipeline, it affects the overall pipeline performance, in terms of prevalence rate reduction and real fix rate. The simulated AI system can obtain the expected theoretical reduction of the prevalence rate, and a final fix rate equal to the theoretical fix rate, only in the best-case scenario, which is when the recall is maximum. In all the other cases, the real reduction of the number of vulnerable code samples, and the final fix rate, can widely vary, falling in the intervals calculated during the simulation. This finding was confirmed when investigating the case study, as it confirms that the final fix rate depends on the oscillation of the classifier recall. 

Second, our simulations show that the uncertainty characterizing the $FN_{ratio}$ is smaller compared to the uncertainty characterizing $\prevalenceaias$ and $\fixaias$. That is, the width of the intervals related to the $FN_{ratio}$ is smaller compared to the intervals of $\prevalenceaias$ and $\fixaias$. However, the incapacity of the pipeline to keep the $FN$ stable between the first and the second classifier could mean overlooking true vulnerabilities due to over-approximation of classifier performance, which could lead to untrustworthy decisions about security risks exposing the possible discrepancy between the preference of risk managers who use the AI system, and the risk tolerance embedded in the system (Paté-Cornell, \citeyear{pate2024preferences}).

The results of the second simulation show that the estimates for the final fix rate are lowered when accounting for shifts in the prevalence rate which can happen when testing and deploying a system, thus demonstrating the importance of accounting for variations in the prevalence rate before deploying a tested system in real-world scenarios. Moreover, the second simulation also shows that increasing the minimum possible recall and specificity that can be sampled has a direct effect on the lower bound of the final fix rate indicating that it is fundamental to understand what is the minimum possible performance of a classifier when employing it in larger AI-augmented systems.

\noindent\textbf{Answer to RQ}
{\emph{How to estimate the total error (or success rate) of the \aias, given the propagating errors of the classifiers in the pipeline?}

Our methodology to assess the risk of propagating uncertainty in a security pipeline can determine the overall intervals for $\prevalenceaias$, $\fixaias$, and $FN_{ratio}$  through simulation. We use it to evaluate the potential propagation of uncertainty on a case study using an AI-based program repair system (AIBUGHUNTER), showing that although the best (claimed) fix rate could be $\fixrate = 47\%$, it could be as low as $3\%$ once uncertainty is accounted for.}

\subsection{Policy implications on AI evaluation}
\label{sec:policy_implications}
The integration of AI sub-systems in safety and security systems will continue, and will progressively align with the evolution of AI models (Collier et al., \citeyear{collier2024good}).

Risk analysis practices are being revolutionized by the integration of AI in several safety and security domains, from cybersecurity (Kaur et al., \citeyear{kaur2023artificial}) to healthcare (Alowais et al., \citeyear{alowais2023revolutionizing}), from predicting natural hazards (Gharehtoragh
\& Johnson, \citeyear{gharehtoragh2024using}) to implementing digital twins (DT), which allow to replicate real-world objects and processes, also in safety and security scenarios (Kreuzer et al., \citeyear{kreuzer2024artificial}). 

As a response to the accompanying risks, new regulations and standardizations have started to come into force worldwide (AI act (European Union, \citeyear{aiact}), the US Executive Order No. 14110 (\citeyear{whitehouse2023executive}), the European Union Aviation Safety Agency (EASA) Artificial Intelligence Roadmap (\citeyear{roadmap2023easa}), the ISO/IEC 42001:2023 (International Standard Organization, \citeyear{isodocument}) and the AI Risk Management Framework (NIST, \citeyear{nistmanagement})).

However, in the process of AI development, application, and regulation, developers, researchers, and policymakers often regard AI models in isolation. They do not consider that AI chains result from the composition of multiple AI models, where the output of one model might become the input for the succeeding model in the toolchain. Even when uncertainty is quantified, uncertainty propagation is ignored, and as our research shows, this can have consequences on the final performance that are elusive to the decision maker. 

In light of our results, we recommend that policies which are being developed to support external and impartial evaluation of AI models should include uncertainty quantification as an explicit indicator. In addition, when systems under evaluation are composed of multiple AI models, the uncertainty quantification should be performed at the system level, quantifying how the uncertainty propagates from one AI model to the next. 

In what follows, we dive into the recently published guidelines on the use of machine learning applications in aviation. By focusing on a concrete safety-critical domain, we highlight the gap regarding 
the quantification of uncertainty propagation and provide recommendations on possible guidelines improvement.

\textbf{Policy recommendations for aviation.}
The necessity to consider uncertainty at the system level has implications for the policies to be adopted in scenarios where AI is applied to safety-risk systems such as in the case of \textit{aviation}.

Although the EASA (\citeyear{roadmap2023easa}; \citeyear{guidance2023easa}), highlights the potential of AI applied to cybersecurity and the importance of uncertainty quantification, a major gap still exists:

\begin{itemize}
    \item Subsystem focus: in the realm of safety assessment and information security, which constitute two important building blocks of the trustworthy AI framework defined by EASA, and of which the first include uncertainty management, the objectives to be reached are characterised at subsystem level (EASA, \citeyear{guidance2023easa}): 

    \begin{displayquote}
        \textbf{Objective SA-01}: The applicant should perform a safety (support) assessment for all AI-based \textbf{(sub)systems}, identifying and addressing specificities introduced by AI/ML usage.
    \end{displayquote}

    \begin{displayquote}
        \textbf{Objective IS-01}: For each AI-based \textbf{(sub)system} and its data sets, the applicant should identify those information security risks with an impact on safety, identifying and addressing specific threats introduced by AI/ML usage.
    \end{displayquote}
\end{itemize}

Contrasting with the EASA approach, our results, related to APR tools but whose implications can be extended also to other \aias s, highlight the importance of modeling uncertainty at the system level, propagating it from the singular subsystems, to verify how the entanglement of the uncertainties of the different components affects the entire system. Thus, to improve the guidelines, we advise integrating the current evaluation policy with additional guidelines emphasizing that safety and risk assessment with the consequent uncertainty quantification, should be performed not only at \textbf{(sub)system} level but also at \textbf{system} level. 

\subsection{Limitations}

\noindent\textbf{No-breaking assumption:}
In our research, we assume that the fixer cannot break the samples that the first classifier classifies as positive when they are negative. Since this is a simplification because we cannot assume that the fixer is perfect and cannot break the code, in future studies we will remove this assumption by experimenting with the breaking-possibility scenario.

\noindent\textbf{No-degradation assumption:}
We assume that all elements that are fixed, cannot be distinguished from \emph{Good} elements from the beginning. The performance of the classifier does not degrade with the fix. We are assuming that the fixer generates code within the same distribution of the originals that are analyzed by the first classifier, thus allowing us to use a second classifier equal to the first. The plan is to use two different classifiers in the future. 

\noindent\textbf{Generalization of simulator to real systems:}
While we assume that the simulation is realistic as it is rooted in relevant theory and recall values reported in related work, we are not working with a real system. In the next step of our research, we will experiment with an actual pipeline, accounting for uncertainty and checking to what extent the results obtained during the simulation are reflected in an actual system. 

\section{Conclusions}
In practice, good performance of APR tools is still challenging to achieve. In a recent publication, Ami et al. (\citeyear{ami2023false}) surveyed 89 practitioners who use automated security testing, and one participant summarized the rate of false positives in reality: \textit{``(At present) 80\% of them are false positives and 20\% of them are something we can fix.''} In addition, the lack of assessing the risk of introducing false negatives into the system is the bigger concern (Ami et al., \citeyear{ami2023false}), which brings challenges for AI-based APR adoption:

\textit{``If the tools miss something, we can not detect that issue, and we just overlook the issues \dots because no one ever reports about false negatives, and we don’t check if the tool ever misses the vulnerabilities''}.

We presented a new approach for assessing the risk of uncertainty propagation and showed, by simulation, that the final performance of an {\aias } may be an entire order of magnitude lower (0.44 vs 0.03) when estimating the effect of propagating errors. Our simulations of the level of uncertainty are in line with the recall values reported in the related work. In addition, the modular implementation of the simulator allows domain experts to use an internal or alternative dataset of recall values, to approximate p-boxes and run a more precise, domain-specific simulation of the propagating uncertainty in their systems. This would allow them to make more informed security risk decisions.

However, future work is needed to validate to what extent the proposed simulation is perceived as useful and how practitioners interpret the communicated uncertainty. For instance, a validation could test whether other factors, connected to real-world and real-time scenarios, such as network traffic and limited bandwidth, or human factors, affect the system's global uncertainty.

Beyond the scenarios modeled in this work, it is worth considering how errors propagate in cases when the fixer modifies a misclassified sample, potentially introducing new vulnerabilities. Moreover, it is worth considering scenarios where the fixer introduces changes with patterns different from the ones that the first classifier is trained to recognize, as it can happen when the classifier and the fixer are trained on different datasets (Fu et al., \citeyear{fu2024aibughunter}), as is often the case, as organizations adopt technologies based on their needs. Capturing these scenarios would allow policymakers to assess when model retraining is required and quantify the drop in residual uncertainty in their systems. 

Finally, improvements in the policies that regulate the evaluation of AI systems are required to guide the risk assessment of AI-based APR tools and in general of AI systems composed of multiple AI models, to quantify the error propagating from (sub)systems to the system level. 

\section*{Acknowledgements}
This work was partially supported by the \textit{Nederlandse Organisatie voor Wetenschappelijk Onderzoek (NWO)} under the KIC HEWSTI Project under grant no. KIC1.VE01.20.004, and the Horizon Europe Sec4AI4Sec Project under grant no. 101120393.

\section*{CRediT statement}
Conceptualization: EM, FM, AP, KT; Methodology: EM, FM; Software: EM, AP; Validation: EM; Formal analysis: EM, FM; Investigation: EM; Resources: na; Data Curation: na; Writing - Original Draft: EM; Writing - Review \& Editing: EM, KT, FM; Visualization: EM; Supervision: FM, KT; Project administration: FM, KT; Funding acquisition: FM, KT; 

\bibliographystyle{plainnat}  

\appendix

\newpage
\section{Monte Carlo (MC) Simulation}
\label{sec:appendix:mc_simulation}
Here we present the results obtained through MC simulation. We implement MC simulation when assuming a constant prevalence rate (Sections \ref{sec:appendix:pr_aias}, \ref{sec:appendix:fr_aias}, \ref{sec:appendix:fn_aias}) and when relaxing this assumption comparing the results with a constant and non-constant prevalence rate (Sections \ref{sec:appendix:second:fr_aias:constant}, \ref{sec:appendix:second:fr_aias:no_constant}). In each section, we present the results when running the MC simulation with 100 sampled recall values and 1000 sampled recall and show, through the standard error of the mean and the percentiles, how the different sample size impacts the precision of the simulation. 

\subsection{Simulation one: constant prevalence rate}

\subsubsection{PR(aias) calculation}
\label{sec:appendix:pr_aias}

\begin{table}[H]
\centering
\caption{The tables show the lower and upper bounds of the final prevalence rate obtained through MC simulation with $100$ sampled recall values ($\mathbf{\prevalenceaias_{100}}$) and $1000$ sampled recall values ($\mathbf{\prevalenceaias_{1000}}$).}
\label{tab:pr_monte_carlo}
\begin{tabular}{@{}lllll@{}}
\hline
\multicolumn{1}{c}{\textbf{}}    & \multicolumn{4}{c}{$\mathbf{\prevalenceaias_{100}}$} \\ \hline
\multicolumn{1}{l|}{$\mathbf{\prevalence}$} & $\mathbf{\fixrate = 0.50}$ & $\mathbf{\fixrate = 0.70}$ & $\mathbf{\fixrate = 0.90}$ & $\mathbf{\fixrate = 1.00}$ \\ \hline
\multicolumn{1}{l|}{0.10}        & {[}0.036, 0.095{]}   & {[}0.016, 0.095{]}   & {[}0.002, 0.093{]}   & {[}0.000, 0.094{]}   \\
\multicolumn{1}{l|}{0.50}        & {[}0.228, 0.488{]}   & {[}0.125, 0.481{]}   & {[}0.033, 0.477{]}   & {[}0.000, 0.472{]}   \\
\multicolumn{1}{l|}{1.00}        & {[}0.463, 0.977{]}   & {[}0.266, 0.972{]}   & {[}0.079, 0.960{]}   & {[}0.000, 0.951{]}   \\ \hline
\end{tabular}
~
\begin{tabular}{@{}lllll@{}}
\hline
\multicolumn{1}{c}{\textbf{}}    & \multicolumn{4}{c}{$\mathbf{\prevalenceaias_{1000}}$} \\ \hline
\multicolumn{1}{l|}{$\mathbf{\prevalence}$} & $\mathbf{\fixrate = 0.50}$ & $\mathbf{\fixrate = 0.70}$ & $\mathbf{\fixrate = 0.90}$ & $\mathbf{\fixrate = 1.00}$ \\ \hline
\multicolumn{1}{l|}{0.10}        & {[}0.049, 0.098{]}   & {[}0.029, 0.097{]}   & {[}0.009, 0.096{]}   & {[}0.000, 0.096{]}   \\
\multicolumn{1}{l|}{0.50}        & {[}0.244, 0.482{]}   & {[}0.145, 0.477{]}   & {[}0.048, 0.471{]}   & {[}0.000, 0.468{]}   \\
\multicolumn{1}{l|}{1.00}        & {[}0.495, 0.972{]}   & {[}0.296, 0.960{]}   & {[}0.097, 0.948{]}   & {[}0.000, 0.943{]}   \\ \hline
\end{tabular}
\end{table}

\begin{table}[H]
\centering
\caption{The tables show the standard error of the mean of the lower and upper bounds of the final prevalence rate, when sampling $100$ recall values ($\mathbf{\sigma_{\prevalenceaias_{100}}}$) and $1000$ recall values ($\mathbf{\sigma_{\prevalenceaias_{1000}}}$).}
\label{tab:pr_monte_carlo_error}
\begin{tabular}{@{}lllll@{}}
\hline
\multicolumn{1}{c}{\textbf{}}    & \multicolumn{4}{c}{$\mathbf{\sigma_{\prevalenceaias_{100}}}$} \\ \hline
\multicolumn{1}{l|}{$\mathbf{\prevalence}$} & $\mathbf{\fixrate = 0.50}$ & $\mathbf{\fixrate = 0.70}$ & $\mathbf{\fixrate = 0.90}$ & $\mathbf{\fixrate = 1.00}$ \\ \hline
\multicolumn{1}{l|}{0.10}        & {[}0.007, 0.014{]}   & {[}0.007, 0.018{]}   & {[}0.008, 0.023{]}   & {[}0.007, 0.026{]}   \\
\multicolumn{1}{l|}{0.50}        & {[}0.004, 0.014{]}   & {[}0.006, 0.019{]}   & {[}0.008, 0.024{]}   & {[}0.008, 0.027{]}   \\
\multicolumn{1}{l|}{1.00}        & {[}0.004, 0.014{]}   & {[}0.005, 0.019{]}   & {[}0.007, 0.025{]}   & {[}0.008, 0.028{]}   \\ \hline
\end{tabular}
~
\begin{tabular}{@{}lllll@{}}
\hline
\multicolumn{1}{c}{\textbf{}}    & \multicolumn{4}{c}{$\mathbf{\sigma_{\prevalenceaias_{1000}}}$} \\ \hline
\multicolumn{1}{l|}{$\mathbf{\prevalence}$} & $\mathbf{\fixrate = 0.50}$ & $\mathbf{\fixrate = 0.70}$ & $\mathbf{\fixrate = 0.90}$ & $\mathbf{\fixrate = 1.00}$ \\ \hline
\multicolumn{1}{l|}{0.10}        & {[}0.001, 0.004{]}   & {[}0.002, 0.006{]}   & {[}0.002, 0.008{]}   & {[}0.002, 0.009{]}   \\
\multicolumn{1}{l|}{0.50}        & {[}0.001, 0.004{]}   & {[}0.002, 0.006{]}   & {[}0.002, 0.007{]}   & {[}0.002, 0.008{]}   \\
\multicolumn{1}{l|}{1.00}        & {[}0.001, 0.004{]}   & {[}0.002, 0.006{]}   & {[}0.002, 0.008{]}   & {[}0.002, 0.008{]}   \\ \hline
\end{tabular}
\end{table}

\begin{table}[H]
\centering
\caption{The tables show the $\nth{25}$ $(P_{25})$, the $\nth{50}$ $(P_{50})$, and the $\nth{75}$ $(P_{75})$ percentiles for the final prevalence rate, when the initial prevalence rate and theoretical fix rate are equal to $0.50$, and when sampling $100$ recall values ($\mathbf{P_{\prevalenceaias_{100}}}$) and $1000$ recall values ($\mathbf{P_{\prevalenceaias_{1000}}}$).}
\label{tab:pr_monte_carlo_percentiles}
\begin{tabular}{llll}
\hline
\multicolumn{4}{c}{$\mathbf{P_{\prevalenceaias_{100}}}$} \\ \hline
$\mathbf{\fixrate}$       & $\mathbf{P_{25}}$ & $\mathbf{P_{50}}$ & $\mathbf{P_{75}}$ \\ \hline
0.50                      & 0.254             & 0.310             & 0.388    \\
0.70                      & 0.156             & 0.233             & 0.338    \\
0.90                      & 0.055             & 0.160             & 0.291    \\
1.00                      & 0.000             & 0.126             & 0.268    \\ \hline
\end{tabular}
~
\begin{tabular}{llll}
\hline
\multicolumn{4}{c}{$\mathbf{P_{\prevalenceaias_{1000}}}$} \\ \hline
$\mathbf{\fixrate}$       & $\mathbf{P_{25}}$ & $\mathbf{P_{50}}$ & $\mathbf{P_{75}}$ \\ \hline
0.50                      & 0.249             & 0.310             & 0.380    \\
0.70                      & 0.149             & 0.236             & 0.334    \\
0.90                      & 0.050             & 0.162             & 0.289    \\
1.00                      & 0.000             & 0.125             & 0.265    \\ \hline
\end{tabular}
\end{table}

\subsubsection{F(aias) calculation}
\label{sec:appendix:fr_aias}

\begin{table}[H]
\caption{The tables show the upper and lower bounds of the final fix rate, when the initial prevalence rate and theoretical fix rate are $0.50$, obtained when sampling $100$ recall values ($\mathbf{\fixaias_{100}}$) and $1000$ recall values ($\mathbf{\fixaias_{1000}}$).}
\centering
\label{tab:fr_monte_carlo}
\begin{tabular}{ll}
\hline
$\mathbf{\fixrate}$ & $\mathbf{\fixaias_{100}}$ \\ \hline
0.50 & {[}0.024, 0.544{]} \\
0.70 & {[}0.038, 0.750{]} \\
0.90 & {[}0.046, 0.934{]} \\
1.00 & {[}0.056, 1.000{]} \\ \hline
\end{tabular}
~
\begin{tabular}{ll}
\hline
$\mathbf{\fixrate}$ & $\mathbf{\fixaias_{1000}}$ \\ \hline
0.50 & {[}0.035, 0.513{]} \\
0.70 & {[}0.047, 0.709{]} \\
0.90 & {[}0.058, 0.904{]} \\
1.00 & {[}0.064, 1.000{]} \\ \hline
\end{tabular}
\end{table}

\begin{table}[H]
\caption{The tables show the standard error of the mean, for the final fix rate, when the initial prevalence rate is $0.50$, with $100$ sampled recall values ($\mathbf{\sigma_{\fixaias_{100}}}$) and $1000$ sampled recall values ($\mathbf{\sigma_{\fixaias_{1000}}}$).}
\centering
\label{tab:fr_monte_carlo_errors}
\begin{tabular}{ll}
\hline
$\mathbf{\fixrate}$ & $\mathbf{\sigma_{\fixaias_{100}}}$ \\ \hline
0.50 & {[}0.004, 0.014{]} \\
0.70 & {[}0.006, 0.019{]} \\
0.90 & {[}0.008, 0.024{]} \\
1.00 & {[}0.008, 0.027{]} \\ \hline
\end{tabular}
~
\begin{tabular}{ll}
\hline
$\mathbf{\fixrate}$ & $\mathbf{\sigma_{\fixaias_{1000}}}$ \\ \hline
0.50 & {[}0.001, 0.004{]} \\
0.70 & {[}0.002, 0.006{]} \\
0.90 & {[}0.002, 0.007{]} \\
1.00 & {[}0.002, 0.008{]} \\ \hline
\end{tabular}
\end{table}

\begin{table}[H]
\centering
\caption{The tables show the $\nth{25}$ $(P_{25})$, the $\nth{50}$ $(P_{50})$, and the $\nth{75}$ $(P_{75})$ percentiles for the final fix rate, when the initial prevalence rate is $0.50$. The tables report the percentiles when sampling $100$ recall values ($\mathbf{P_{\fixaias_{100}}}$) and $1000$ recall values ($\mathbf{P_{\fixaias_{1000}}}$).}
\label{tab:fr_monte_carlo_percentiles}
\begin{tabular}{llll}
\hline
\multicolumn{4}{c}{$\mathbf{P_{\fixaias_{100}}}$} \\ \hline
$\mathbf{\fixrate}$       & $\mathbf{P_{25}}$ & $\mathbf{P_{50}}$ & $\mathbf{P_{75}}$ \\ \hline
0.50                      & 0.223             & 0.380             & 0.492    \\
0.70                      & 0.325             & 0.534             & 0.688    \\
0.90                      & 0.418             & 0.681             & 0.891    \\
1.00                      & 0.464             & 0.747             & 1.000    \\ \hline
\end{tabular}
~
\begin{tabular}{llll}
\hline
\multicolumn{4}{c}{$\mathbf{P_{\fixaias_{1000}}}$} \\ \hline
$\mathbf{\fixrate}$       & $\mathbf{P_{25}}$ & $\mathbf{P_{50}}$ & $\mathbf{P_{75}}$ \\ \hline
0.50                      & 0.239             & 0.380             & 0.503    \\
0.70                      & 0.332             & 0.528             & 0.701    \\
0.90                      & 0.423             & 0.677             & 0.900    \\
1.00                      & 0.469             & 0.750             & 1.000    \\ \hline
\end{tabular}
\end{table}

\subsubsection{$\mathbf{FN_{ratio}}$ calculation}
\label{sec:appendix:fn_aias}

\begin{table}[H]
\caption{The tables show the upper and lower bounds for the false negatives ratio, when sampling 100 recall values ($\mathbf{{FN_{ratio}}_{100}}$) and 1000 recall values ($\mathbf{{FN_{ratio}}_{1000}}$).}
\centering
\label{tab:fn_monte_carlo}
\begin{tabular}{ll}
\hline
$\mathbf{\fixrate}$ & $\mathbf{{FN_{ratio}}_{100}}$ \\ \hline
0.50 & {[}1.111, 1.929{]} \\
0.70 & {[}1.087, 1.389{]} \\
0.90 & {[}1.160, 1.176{]} \\
1.00 & {[}1.000, 1.000{]} \\ \hline
\end{tabular}
~
\begin{tabular}{ll}
\hline
$\mathbf{\fixrate}$ & $\mathbf{{FN_{ratio}}_{1000}}$\\ \hline
0.50 & {[}1.376, 1.403{]} \\
0.70 & {[}1.221, 1.236{]} \\
0.90 & {[}1.081, 1.075{]} \\
1.00 & {[}1.000, 1.000{]} \\ \hline
\end{tabular}
\end{table}

\begin{table}[H]
\caption{The tables show the standard error of the mean of the lower and upper bounds of the final false negatives ratio when sampling 100 recall values ($\mathbf{{\sigma_{FN_{ratio}}}_{100}}$) and 1000 recall values ($\mathbf{{\sigma_{FN_{ratio}}}_{1000}}$).}
\centering
\label{tab:fn_monte_carlo_errors}
\begin{tabular}{ll}
\hline
$\mathbf{\fixrate}$ & $\mathbf{{\sigma_{FN_{ratio}}}_{100}}$ \\ \hline
0.50 & {[}0.023, 0.018{]} \\
0.70 & {[}0.019, 0.010{]} \\
0.90 & {[}0.006, 0.005{]} \\
1.00 & {[}0.000, 0.000{]} \\ \hline
\end{tabular}
~
\begin{tabular}{ll}
\hline
$\mathbf{\fixrate}$ & $\mathbf{{\sigma_{FN_{ratio}}}_{1000}}$ \\ \hline
0.50 & {[}0.006, 0.004{]} \\
0.70 & {[}0.004, 0.003{]} \\
0.90 & {[}0.001, 0.001{]} \\
1.00 & {[}0.000, 0.000{]} \\ \hline
\end{tabular}
\end{table}

\begin{table}[H]
\centering
\caption{The tables show the $\nth{25}$ $(P_{25})$, the $\nth{50}$ $(P_{50})$, and the $\nth{75}$ $(P_{75})$ percentiles for the final $FN_{ratio}$. The tables report the percentiles when sampling $100$ recall values ($\mathbf{{P_{FN_{ratio}}}_{100}}$) and $1000$ recall values ($\mathbf{{P_{FN_{ratio}}}_{1000}}$).}
\label{tab:fn_monte_carlo_percentiles}
\begin{tabular}{llll}
\hline
\multicolumn{4}{c}{$\mathbf{{P_{FN_{ratio}}}_{100}}$} \\ \hline
$\mathbf{\fixrate}$       & $\mathbf{P_{25}}$ & $\mathbf{P_{50}}$ & $\mathbf{P_{75}}$ \\ \hline
0.50                      & 1.315             & 1.520             & 1.724    \\
0.70                      & 1.162             & 1.238             & 1.313    \\
0.90                      & 1.164             & 1.168             & 1.172    \\
1.00                      & 1.000             & 1.000             & 1.000    \\ \hline
\end{tabular}
~
\begin{tabular}{llll}
\hline
\multicolumn{4}{c}{$\mathbf{{P_{FN_{ratio}}}_{1000}}$} \\ \hline
$\mathbf{\fixrate}$       & $\mathbf{P_{25}}$ & $\mathbf{P_{50}}$ & $\mathbf{P_{75}}$ \\ \hline
0.50                      & 1.382             & 1.389             & 1.395    \\
0.70                      & 1.225             & 1.229             & 1.233    \\
0.90                      & 1.076             & 1.078             & 1.079    \\
1.00                      & 1.000             & 1.000             & 1.000    \\ \hline
\end{tabular}
\end{table}

\subsection{Simulation two: beyond constant prevalence rate}

\subsubsection{F(aias) calculation with constant prevalence rate}
\label{sec:appendix:second:fr_aias:constant}

\begin{table}[H]
\centering
\caption{The tables show the final fix rate calculated when the prevalence rate is constant, respectively when the number of sampled recall values is 100 ($\mathbf{\fixaias_{100}}$) and when the number of sampled recall values is 1000 ($\mathbf{\fixaias_{1000}}$).}
\label{tab:const_prev_analysis_2_mc}
\begin{tabular}{@{}llll@{}}
\hline
 & \multicolumn{3}{c}{\textbf{$\mathbf{\fixaias_{100}}$}} \\ \hline
\multicolumn{1}{c}{\textbf{}} & \multicolumn{3}{c}{\textbf{Min. Rec and Spec}} \\ \hline
\multicolumn{1}{c|}{\textbf{$\mathbf{\fixrate}$}} & \textbf{Min = 0.50} & \textbf{Min = 0.70} & \textbf{Min = 0.90} \\ \hline
\multicolumn{1}{l|}{0.50} & {[}0.240, 0.690{]} & {[}0.350, 0.680{]} & {[}0.450, 0.710{]} \\
\multicolumn{1}{l|}{0.70} & {[}0.320, 0.850{]} & {[}0.440, 0.840{]} & {[}0.570, 0.850{]} \\
\multicolumn{1}{l|}{0.90} & {[}0.410, 0.980{]} & {[}0.600, 0.970{]} & {[}0.750, 0.970{]} \\
\multicolumn{1}{l|}{1.00} & {[}0.470, 1.000{]} & {[}0.640, 1.000{]} & {[}0.840, 1.000{]} \\ \hline
\end{tabular}
~
\begin{tabular}{@{}llll@{}}
\hline
 & \multicolumn{3}{c}{\textbf{$\mathbf{\fixaias_{1000}}$}} \\ \hline
\multicolumn{1}{c}{\textbf{}} & \multicolumn{3}{c}{\textbf{Min. Rec and Spec}} \\ \hline
\multicolumn{1}{c|}{\textbf{$\mathbf{\fixrate}$}} & \textbf{Min = 0.50} & \textbf{Min = 0.70} & \textbf{Min = 0.90} \\ \hline
\multicolumn{1}{l|}{0.50} & {[}0.239, 0.519{]} & {[}0.340, 0.517{]} & {[}0.436, 0.516{]} \\
\multicolumn{1}{l|}{0.70} & {[}0.338, 0.717{]} & {[}0.476, 0.717{]} & {[}0.618, 0.713{]} \\
\multicolumn{1}{l|}{0.90} & {[}0.434, 0.912{]} & {[}0.618, 0.910{]} & {[}0.798, 0.910{]} \\
\multicolumn{1}{l|}{1.00} & {[}0.489, 1.000{]} & {[}0.688, 1.000{]} & {[}0.891, 1.000{]} \\ \hline
\end{tabular}
\end{table}

\begin{table}[H]
\centering
\caption{The tables show the standard error of the mean for the upper and lower bound of the final fix rate, with constant prevalence rate, and when the number of sampled recall values is $100$ ($\mathbf{{\sigma_{\fixaias}}_{100}}$) and the number of sampled recall values is $1000$ ($\mathbf{{\sigma_{\fixaias}}_{1000}}$)}
\label{tab:const_prev_analysis_mc_errors}
\begin{tabular}{@{}llll@{}}
\hline
 & \multicolumn{3}{c}{$\mathbf{{\sigma_{\fixaias}}_{100}}$} \\ \hline
\multicolumn{1}{c}{\textbf{}} & \multicolumn{3}{c}{\textbf{Min. Rec and Spec}} \\ \hline
\multicolumn{1}{c|}{\textbf{$\mathbf{\fixrate}$}} & \textbf{Min = 0.50} & \textbf{Min = 0.70} & \textbf{Min = 0.90} \\ \hline
\multicolumn{1}{l|}{0.50} & {[}0.006, 0.006{]} & {[}0.005, 0.005{]} & {[}0.005, 0.004{]} \\
\multicolumn{1}{l|}{0.70} & {[}0.007, 0.007{]} & {[}0.005, 0.006{]} & {[}0.005, 0.005{]} \\
\multicolumn{1}{l|}{0.90} & {[}0.008, 0.008{]} & {[}0.005, 0.006{]} & {[}0.003, 0.003{]} \\
\multicolumn{1}{l|}{1.00} & {[}0.008, 0.009{]} & {[}0.005, 0.007{]} & {[}0.002, 0.003{]} \\ \hline
\end{tabular}
~
\begin{tabular}{@{}llll@{}}
\hline
 & \multicolumn{3}{c}{$\mathbf{{\sigma_{\fixaias}}_{1000}}$} \\ \hline
\multicolumn{1}{c}{\textbf{}} & \multicolumn{3}{c}{\textbf{Min. Rec and Spec}} \\ \hline
\multicolumn{1}{c|}{\textbf{$\mathbf{\fixrate}$}} & \textbf{Min = 0.50} & \textbf{Min = 0.70} & \textbf{Min = 0.90} \\ \hline
\multicolumn{1}{l|}{0.50} & {[}0.001, 0.001{]} & {[}0.001, 0.001{]} & {[}0.000, 0.000{]} \\
\multicolumn{1}{l|}{0.70} & {[}0.002, 0.002{]} & {[}0.001, 0.001{]} & {[}0.000, 0.000{]} \\
\multicolumn{1}{l|}{0.90} & {[}0.003, 0.003{]} & {[}0.002, 0.002{]} & {[}0.001, 0.001{]} \\
\multicolumn{1}{l|}{1.00} & {[}0.003, 0.003{]} & {[}0.002, 0.002{]} & {[}0.001, 0.001{]} \\ \hline
\end{tabular}
\end{table}

\begin{table}[H]
\centering
\caption{The tables show the $\nth{25}$ $(P_{25})$, the $\nth{50}$ $(P_{50})$, and the $\nth{75}$ $(P_{75})$ percentiles for the final fix rate when prevalence rate is constant, the minimum recall is equal to 0.50. The tables report the percentiles when sampling $100$ recall values ($\mathbf{{P_{\fixaias}}_{100}}$) and $1000$ recall values ($\mathbf{{P_{\fixaias}}_{1000}}$).}
\label{tab:fr_monte_carlo_percentiles_constant}
\begin{tabular}{llll}
\hline
\multicolumn{4}{c}{$\mathbf{{P_{\fixaias}}_{100}}$} \\ \hline
$\mathbf{\fixrate}$       & $\mathbf{P_{25}}$ & $\mathbf{P_{50}}$ & $\mathbf{P_{75}}$ \\ \hline
0.50                      & 0.378             & 0.460             & 0.540    \\
0.70                      & 0.470             & 0.590             & 0.700    \\
0.90                      & 0.570             & 0.715             & 0.890    \\
1.00                      & 0.618             & 0.770             & 1.000    \\ \hline
\end{tabular}
~
\begin{tabular}{llll}
\hline
\multicolumn{4}{c}{$\mathbf{{P_{\fixaias}}_{1000}}$} \\ \hline
$\mathbf{\fixrate}$       & $\mathbf{P_{25}}$ & $\mathbf{P_{50}}$ & $\mathbf{P_{75}}$ \\ \hline
0.50                      & 0.260             & 0.377             & 0.492    \\
0.70                      & 0.360             & 0.526             & 0.687    \\
0.90                      & 0.461             & 0.676             & 0.881    \\
1.00                      & 0.511             & 0.750             & 0.981    \\ \hline
\end{tabular}
\end{table}

\subsubsection{F(aias) calculation without constant prevalence rate}
\label{sec:appendix:second:fr_aias:no_constant}

\begin{table}[H]
\centering
\caption{The tables show the calculation of the upper and lower bounds of the final fix rate without a constant prevalence rate. The two tables show the results when the number of recall values sampled is 100 ($\mathbf{\fixaias_{100}}$) and when the number of recall values sampled is 1000 ($\mathbf{\fixaias_{1000}}$).}
\label{tab:non_cost_prev_analysis_2_mc}
\begin{tabular}{@{}llll@{}}
\hline
 & \multicolumn{3}{c}{$\mathbf{\fixaias_{100}}$} \\ \hline
\multicolumn{1}{c}{\textbf{}} & \multicolumn{3}{c}{\textbf{Min. Rec and Spec}} \\ \hline
\multicolumn{1}{c|}{\textbf{$\mathbf{\fixrate}$}} & \textbf{min = 0.50} & \textbf{min = 0.70} & \textbf{min = 0.90} \\ \hline
\multicolumn{1}{l|}{0.50} & {[}0.000, 0.690{]} & {[}0.319, 0.690{]} & {[}0.640, 0.710{]} \\
\multicolumn{1}{l|}{0.70} & {[}0.000, 0.850{]} & {[}0.190, 0.863{]} & {[}0.667, 0.850{]} \\
\multicolumn{1}{l|}{0.90} & {[}0.000, 0.900{]} & {[}0.250, 0.875{]} & {[}0.700, 0.950{]} \\
\multicolumn{1}{l|}{1.00} & {[}0.000, 0.910{]} & {[}0.233, 0.921{]} & {[}0.733, 0.963{]} \\ \hline
\end{tabular}
~
\begin{tabular}{@{}llll@{}}
\hline
 & \multicolumn{3}{c}{$\mathbf{\fixaias_{1000}}$} \\ \hline
\multicolumn{1}{c}{\textbf{}} & \multicolumn{3}{c}{\textbf{Min. Rec and Spec}} \\ \hline
\multicolumn{1}{c|}{\textbf{$\mathbf{\fixrate}$}} & \textbf{min = 0.50} & \textbf{min = 0.70} & \textbf{min = 0.90} \\ \hline
\multicolumn{1}{l|}{0.50} & {[}0.000, 0.519{]} & {[}0.340, 0.519{]} & {[}0.550, 0.630{]} \\
\multicolumn{1}{l|}{0.70} & {[}0.000, 0.717{]} & {[}0.345, 0.717{]} & {[}0.650, 0.713{]} \\
\multicolumn{1}{l|}{0.90} & {[}0.000, 0.770{]} & {[}0.460, 0.790{]} & {[}0.756, 0.860{]} \\
\multicolumn{1}{l|}{1.00} & {[}0.000, 0.771{]} & {[}0.550, 0.810{]} & {[}0.851, 0.920{]} \\ \hline
\end{tabular}
\end{table}

\begin{table}[H]
\centering
\caption{The tables show the standard error of the mean for the lower and upper bounds of the final fix rate when the initial prevalence rate is not constant. The two tables show the results when the number of recall values is 100 ($\mathbf{{\sigma_{\fixaias}}_{100}}$) and when the number of recall values is 1000 ($\mathbf{{\sigma_{\fixaias}}_{1000}}$).}
\label{tab:non_cost_prev_analysis_mc_errors}
\begin{tabular}{@{}llll@{}}
\hline
 & \multicolumn{3}{c}{$\mathbf{{\sigma_{\fixaias}}_{100}}$} \\ \hline
\multicolumn{1}{c}{\textbf{}} & \multicolumn{3}{c}{\textbf{Min. Rec and Spec}} \\ \hline
\multicolumn{1}{c|}{\textbf{$\mathbf{\fixrate}$}} & \textbf{min = 0.50} & \textbf{min = 0.70} & \textbf{min = 0.90} \\ \hline
\multicolumn{1}{l|}{0.50} & {[}0.005, 0.016{]} & {[}0.005, 0.008{]} & {[}0.004, 0.001{]} \\
\multicolumn{1}{l|}{0.70} & {[}0.008, 0.016{]} & {[}0.003, 0.009{]} & {[}0.001, 0.002{]} \\
\multicolumn{1}{l|}{0.90} & {[}0.015, 0.017{]} & {[}0.008, 0.010{]} & {[}0.002, 0.003{]} \\
\multicolumn{1}{l|}{1.00} & {[}0.019, 0.017{]} & {[}0.011, 0.011{]} & {[}0.004, 0.004{]} \\ \hline
\end{tabular}
~
\begin{tabular}{@{}llll@{}}
\hline
 & \multicolumn{3}{c}{$\mathbf{{\sigma_{\fixaias}}_{1000}}$} \\ \hline
\multicolumn{1}{c}{\textbf{}} & \multicolumn{3}{c}{\textbf{Min. Rec and Spec}} \\ \hline
\multicolumn{1}{c|}{\textbf{$\mathbf{\fixrate}$}} & \textbf{min = 0.50} & \textbf{min = 0.70} & \textbf{min = 0.90} \\ \hline
\multicolumn{1}{l|}{0.50} & {[}0.001, 0.005{]} & {[}0.002, 0.003{]} & {[}0.001, 0.000{]} \\
\multicolumn{1}{l|}{0.70} & {[}0.003, 0.005{]} & {[}0.001, 0.003{]} & {[}0.000, 0.000{]} \\
\multicolumn{1}{l|}{0.90} & {[}0.005, 0.006{]} & {[}0.003, 0.003{]} & {[}0.001, 0.001{]} \\
\multicolumn{1}{l|}{1.00} & {[}0.006, 0.006{]} & {[}0.004, 0.004{]} & {[}0.001, 0.001{]} \\ \hline
\end{tabular}
\end{table}

\begin{table}[H]
\centering
\caption{The tables show the $\nth{25}$ $(P_{25})$, the $\nth{50}$ $(P_{50})$, and the $\nth{75}$ $(P_{75})$ percentiles for the final fix rate when prevalence rate is not constant and when minimum recall and specificity are equal to $0.50$. The tables report the percentiles when sampling 100 recall values ($\mathbf{{P_{\fixaias}}_{100}}$) and 1000 recall values ($\mathbf{{P_{\fixaias}}_{1000}}$).}
\label{tab:fr_monte_carlo_percentiles_non_constant}
\begin{tabular}{llll}
\hline
\multicolumn{4}{c}{$\mathbf{{P_{\fixaias}}_{100}}$} \\ \hline
$\mathbf{\fixrate}$       & $\mathbf{P_{25}}$ & $\mathbf{P_{50}}$ & $\mathbf{P_{75}}$ \\ \hline
0.50                      & 0.067             & 0.440             & 0.571    \\
0.70                      & 0.064             & 0.461             & 0.711    \\
0.90                      & 0.062             & 0.456             & 0.853    \\
1.00                      & 0.078             & 0.459             & 0.891    \\ \hline
\end{tabular}
~
\begin{tabular}{llll}
\hline
\multicolumn{4}{c}{$\mathbf{{P_{\fixaias}}_{1000}}$} \\ \hline
$\mathbf{\fixrate}$       & $\mathbf{P_{25}}$ & $\mathbf{P_{50}}$ & $\mathbf{P_{75}}$ \\ \hline
0.50                      & 0.009             & 0.440             & 0.505    \\
0.70                      & 0.008             & 0.458             & 0.697    \\
0.90                      & 0.009             & 0.475             & 0.731    \\
1.00                      & 0.009             & 0.481             & 0.747    \\ \hline
\end{tabular}
\end{table}

\section{Formula derivations}

\subsection{Derivation of \aias \@\xspace fix rate from the positives}
\label{sec:pos_aias_derivation}
\begin{align}
    \positives(aias) & = (1-\fixrate)\cdot TP_{\nth{1}} + FN_{\nth{1}}  \\
    \positives(aias) & = \positives - \fixaias \cdot \positives \\     \prevalence \cdot \total - \fixaias \cdot \prevalence \cdot \total
    & =  (1-\fixrate) \cdot \recall \cdot \prevalence \cdot \total + (1 - \recall) \cdot \prevalence \cdot \total \\
    1 - \fixaias 
    & =  (1-\fixrate) \cdot \recall + (1 - \recall) \\
    \fixaias 
    & =  1 -  (1-\fixrate) \cdot \recall - (1 - \recall) \\
    & =  1 -  \recall +\fixrate \cdot \recall - (1 - \recall) \\
    & =  \fixrate \cdot \recall 
 \end{align}

\subsection{Derivation of $\mathbf{\prevalence(aias)}$ from $\mathbf{Pos(aias)}$}
\label{sec:pr_aias_derivation}
\begin{align}
    \positives(aias) & = \positives - \fixaias \cdot \positives\\
    \prevalence(aias) \cdot \total & = \prevalence \cdot \total - \fixrate \cdot\recall \cdot \prevalence \cdot \total \\
    \prevalence(aias)  & = (1- \fixrate \cdot\recall) \cdot \prevalence 
\end{align}

\subsection{Derivation of $\mathbf{TPR(aias)}$}
\label{sec:tpr_aias_derivation}
\begin{align}
    TPR(aias) &=\frac{TP_{\nth{2}}}{\positives(aias)} \\
    & = \frac{(1-\fixrate)\cdot TP_{\nth{1}}\cdot \recall}{\positives - \fixaias \cdot \positives} \\
    & = \frac{(1-\fixrate)\cdot \recall\cdot \prevalence \cdot \total\cdot \recall}{\prevalence \cdot \total - \fixrate\cdot \recall \cdot \prevalence \cdot \total} \\
    & = \frac{(1-\fixrate)\cdot \recall\cdot \recall}{1 - \fixrate\cdot \recall }
\end{align}

\subsection{$\mathbf{TPR(aias)} \leq \mathbf{TPR}$}
\label{sec:tpr_aias_tpr}
\begin{align}
    \frac{(1-\fixrate)\cdot \recall}{1 - \fixrate\cdot\recall } & \leq 1\\
    (1-\fixrate)\cdot \recall & \leq 1 - \fixrate\cdot\recall\\
    \recall-\fixrate\cdot \recall & \geq 1 - \fixrate\cdot\recall\\
    \recall & \geq 1    
\end{align} 

\subsection{Derivation of the false positives}
\label{sec:fp_aias_derivation}
\begin{align}
     FP(aias) &= \recall \cdot \frac{1- \precision}{\precision} \cdot (1-\fixrate)\cdot \recall \cdot \prevalence \cdot \total 
\end{align}

\subsection{Derivation of the $\mathbf{FAR(aias)}$}
\label{sec:far_aias_derivation}
\begin{align}
       FAR(aias) &= \frac{FP(aias)}{\negatives(aias)} = \frac{FP(aias)}{\total-\positives(aias)} \\
                & = \frac{\recall \cdot \frac{1- \precision}{\precision} \cdot (1-\fixrate)\cdot \recall \cdot \prevalence \cdot \total}{\total - (\prevalence \cdot \total - \fixaias \cdot \prevalence \cdot \total) }  \\ 
                & = \frac{\recall \cdot \frac{1 - \precision}{\precision} \cdot (1-\fixrate)\cdot \recall \cdot \prevalence }{1 - (\prevalence - \fixrate\cdot\recall \cdot \prevalence) }  \\ 
                & = \recall \cdot \frac{1- \precision}{\precision}\frac{(1-\fixrate)\cdot \recall \cdot \prevalence}{1 - (1 - \fixrate\cdot\recall )\cdot \prevalence}  \\ 
                & = \recall^2 \cdot \frac{1- \precision}{\precision}\frac{(1-\fixrate)\cdot\prevalence}{1 - (1 - \fixrate\cdot\recall)\cdot \prevalence}
\end{align}

\subsection{Proof that the \aias \xspace false alert rate is less than or equal to the false alert rate of the first classifier ($\mathbf{FAR(aias) \leq FAR}$)}
\label{sec:far_aias_less_far}

\begin{align}
FAR({aias}) & \leq FAR \\
        \recall^2 \cdot \frac{1- \precision}{\precision}\frac{(1-\fixrate)\cdot\prevalence }{1 - (1 - \fixrate\cdot\recall)\cdot \prevalence} & \leq \recall \cdot \frac{1- \precision}{\precision} \frac{\prevalence\cdot \total}{\total-\prevalence\cdot\total} \\
        \recall\cdot \frac{(1-\fixrate)\cdot\prevalence }{1 - (1 - \fixrate\cdot\recall)\cdot \prevalence} & \leq \frac{\prevalence}{1-\prevalence} \\
        \recall\cdot \frac{(1-\fixrate)}{1 - (1 - \fixrate\cdot\recall)\cdot \prevalence} & \leq \frac{1}{1-\prevalence} \\
        \recall\cdot (1-\fixrate)(1-\prevalence) & \leq 1 - (1 - \fixrate\cdot\recall)\cdot \prevalence \\
        \recall\cdot (1-\fixrate-\prevalence+\fixrate\cdot\prevalence) & \leq 1 -\prevalence  + \fixrate\cdot\recall\cdot\prevalence \\
        \recall-\recall\cdot\fixrate-\recall\cdot\prevalence+\recall\cdot\fixrate\cdot\prevalence & \leq 1 -\prevalence  + \fixrate\cdot\recall\cdot\prevalence \\
        \recall-\recall\cdot\fixrate-\recall\cdot\prevalence & \leq 1 -\prevalence  \\
        \recall\cdot(1-\fixrate-\prevalence) & \leq 1 -\prevalence  \\
        \ \ \mbox{ if }  1-\fixrate-\prevalence > 0 \mbox{ which is } 1 > \fixrate+\prevalence \\
        \recall & \leq \frac{1 -\prevalence}{1-\fixrate-\prevalence}    \mbox{ and } 1-\fixrate-\prevalence \leq 1-\prevalence \mbox{ implies }
        1 \leq \frac{1 -\prevalence}{1-\fixrate-\prevalence} \\
        \recall & \leq 1 \leq \frac{1 -\prevalence}{1-\fixrate-\prevalence} \mbox{ always true } \\
        \mbox{ if }  1-\fixrate-\prevalence < 0  \mbox{ which is } 1< \fixrate+\prevalence\\
        \recall & \geq \frac{\prevalence-1}{\fixrate+\prevalence-1} \mbox{ and } \prevalence-1 \geq 0 \mbox{ implies } \frac{\prevalence-1}{\fixrate+\prevalence-1} \geq 0 \\
         \recall & \geq 0 \geq \frac{\prevalence-1}{\fixrate+\prevalence-1} \mbox{ always true }
\end{align}

\subsection{Derivation of the total number of elements passed to the fixer}
\label{sec:n2_derivation}
\begin{align}
     \total_{\nth{2}} &= TP_{\nth{1}} + FP_{\nth{1}} \\
                 & = \recall \cdot \prevalence \cdot \total + \recall \cdot \frac{1 - \precision}{\precision} \cdot \prevalence \cdot \total \\
                 &= \frac{\precision \cdot \recall  + \recall - \recall \cdot \precision}{\precision} \cdot \prevalence \cdot \total\\
                 &= \frac{\recall}{\precision} \cdot \prevalence \cdot \total
\end{align}

\subsection{Derivation of the false positives starting from the precision}
\label{sec_fp_derivation}
\begin{align}
    \precision & = \frac{TP}{TP + FP} \\
   (TP + FP) \cdot \precision & = TP \\
   FP \cdot \precision & = TP \cdot (1 - \precision) \\
   FP & = \positives \cdot \recall \cdot \frac{1 - \precision}{\precision}
\end{align}

\subsection{Derivation of the final number of true positives}
\label{sec:tp_derivation}
\begin{align}
    TP(aias) & = (1 - \fixrate) \cdot TP_{\nth{1}} \cdot \recall \\
             & = (1 - \fixrate) \cdot (\positives \cdot \recall) \cdot \recall \\
             & = (1 - \fixrate) \cdot \recall^2 \cdot \positives \\
             & = (1 - \fixrate) \cdot \recall^2 \cdot \prevalence \cdot \total
\end{align}

\subsection{Derivation of the \aias \xspace false negatives ($\mathbf{FN(aias)}$)}
\label{sec:fn_derivation}
\begin{align}
     FN(aias) &= \left[(1 - \fixrate) \cdot TP_{\nth{1}}\right] \cdot (1 - \recall) + FN_{\nth{1}} \\
             &= \left[(1 - \fixrate) \cdot (\positives \cdot \recall)\right] \cdot (1 - \recall) + (\positives \cdot (1 - \recall))  \\ 
             &= \left\{\left[(1 - \fixrate) \cdot \recall\right] \cdot (1 - \recall) +  (1 - \recall) \right\} \cdot \positives \\ 
             &= \left\{\left[(1 - \fixrate) \cdot \recall\right] + 1 \right\} \cdot (1 - \recall) \cdot \positives \\ 
             &= \left[1+(1 - \fixrate) \cdot \recall\right] \cdot (1 - \recall) \cdot \prevalence \cdot \total \\ 
             &= \left[1+(1 - \fixrate) \cdot \recall\right] \cdot FN_{\nth{1}}
\end{align}

\subsection{Derivation of the \aias \xspace prevalence rate ($\mathbf{\prevalenceaias}$)}
\label{sec:pr_end_derivation}
\begin{align}
    \prevalence(aias) &= \frac{TP(aias) + FN(aias)}{TP(aias) + FN(aias) + TN(aias) + FP(aias)} \\
    & = \frac{TP(aias) + FN(aias)}{\total} \\
    & = \frac{(1 - \fixrate) \cdot \recall^2 \cdot \prevalence \cdot \total + \left[1+(1 - \fixrate)\cdot \recall\right]\cdot (1 - \recall) \cdot \prevalence \cdot \total}{\total} \\
    & = (1 - \fixrate) \cdot \recall^2 \cdot \prevalence + \left[1+(1 - \fixrate)\cdot \recall\right]\cdot (1 - \recall) \cdot \prevalence \\
    & = \left[(1 - \fixrate) \cdot \recall^2 + \left[1+(1 - \fixrate)\cdot \recall\right]\cdot (1 - \recall) \right]\cdot \prevalence \\
    & = \left[(1 - \fixrate) \cdot \recall^2 + (1 - \recall) + (1 - \fixrate)\cdot \recall\cdot(1 - \recall)\right]\cdot \prevalence\\
    & = \left[(1 - \fixrate) \cdot \recall^2 + (1 - \recall) + (1 - \fixrate)\cdot \recall - (1 - \fixrate)\cdot \recall^2)\right]\cdot \prevalence\\
    & = \left[1 - \recall + \recall - \fixrate \cdot \recall)\right]\cdot \prevalence \\
    & = \left[1 - \fixrate \cdot \recall\right]\cdot \prevalence
\end{align}

\end{document}